\documentclass[
    aps,
    prx,
    reprint,
    superscriptaddress,
    amsmath,
    amssymb,
    nofootinbib,
    longbibliography,
    floatfix
]{revtex4-2} 

\usepackage{bm}
\usepackage{braket}
\usepackage[usenames,dvipsnames, pdftex]{xcolor}
\usepackage[colorlinks=true,
            linkcolor=blue,
            citecolor=blue,
            urlcolor=blue]{hyperref}
\usepackage[caption=false]{subfig}
\usepackage{natbib}
\usepackage{adjustbox}
\usepackage{graphicx}

\newcommand{\Tr}{\operatorname{Tr}}
\newcommand{\ms}{\mathcal{S}}
\newcommand{\stab}{\text{STAB}_{n}}

\begin{document}
 
\title{Such stuff as magic is made on: compact operator algebra, 
stabilizer polytope and the structure of reduced density matrices 
in a Kitaev spin liquid}

\author{Snigdh Sabharwal}
\email{snigdh.sabharwal@oist.jp}
\affiliation{Theory of Quantum Matter Unit, Okinawa Institute of Science and Technology Graduate University, Onna-son, Okinawa 904-0412, Japan}

\author{Nic Shannon}
\affiliation{Theory of Quantum Matter Unit, Okinawa Institute of Science and Technology Graduate University, Onna-son, Okinawa 904-0412, Japan}

\author{Paul Skrzypczyk}
\affiliation{H. H. Wills Physics Laboratory, University of Bristol, Tyndall Avenue, Bristol, BS8 1TL, UK.}
\date{\today}

\begin{abstract}

Quantum many-body systems exhibit rich and complex behaviour. Magic has recently emerged as a powerful new diagnostic for probing such systems, complementing other key features such as many-body entanglement. Here, starting from an investigation of the onset of magic within subsystems of a larger many-body quantum system, we show that intricate structures emerge which shed important light on the underlying many-body physics. Focusing on the Kitaev honeycomb model, we first identify the temperature below which local subsystems acquire magic. Remarkably, we find that the optimal magic witnesses active at the onset of magic reveal compact operator spaces that continue to capture the local thermal states throughout their subsequent evolution. For the six-site hexagonal marginal, this connection can be made stronger: the same operator space emerges independently from the symmetries of the local marginal and forms the symmetry-resolved plaquette algebra. When these symmetries are realized exactly, the local state lies entirely within this algebra and the reduced robustness of magic is equal to the full robustness of magic. The same reduced description can also be applied to other local quantum resources, which we illustrate using genuine multipartite entanglement. Furthermore, the operator spaces possess a rich algebraic structure in the form of finite-dimensional Euclidean Jordan algebras, with a natural interpretation in terms of bond and bond-cycle operators. Our general methodology therefore show how the onset of local magic can reveal a compact, physically meaningful operator structure underlying the finite-temperature Kitaev spin liquid, and opens up a new avenue towards reduced descriptions of quantum resources in many-body systems.
\end{abstract}

\maketitle

\section{Introduction}
\begin{figure*}[t!]
    \centering
    \includegraphics[width=1.0\linewidth]{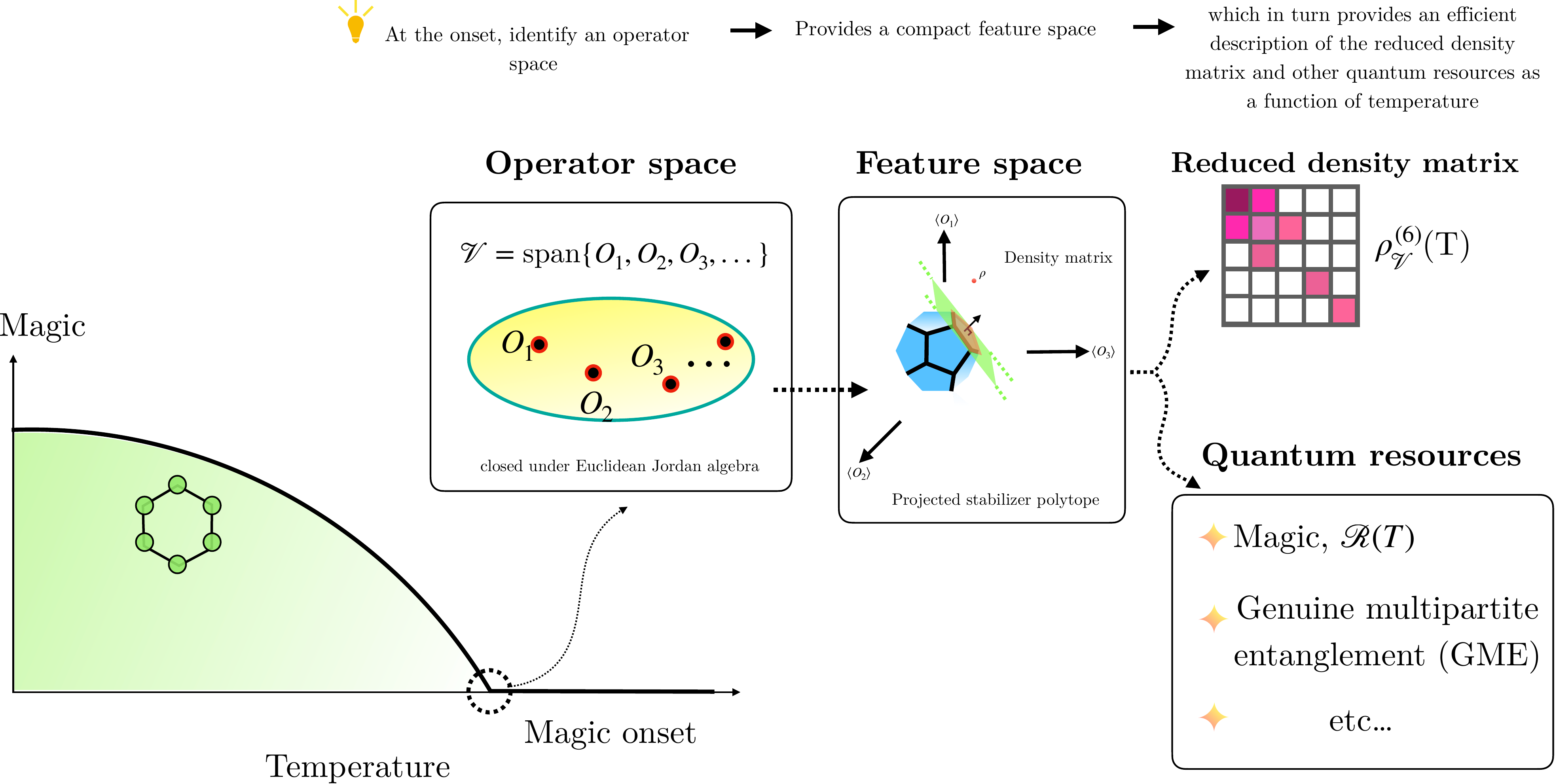}
    \caption{A schematic illustration of how the magic witness structure at the onset temperature provides a reduced operator space using which one can obtain a compact representation of the stabilizer polytope, an efficient representation of the reduced density matrix and other quantum resources such as genuine multipartite entanglement, across the entire temperature range.}
    \label{fig:overview.schematic}
\end{figure*}

Entanglement has long served as a central diagnostic to characterize quantum many-body phases \cite{Amico2008,Laflorencie2016quantum}. However, entanglement alone does not exhaust the quantum resources present in a many-body system. Stabilizer states, for example, can be highly entangled and yet admit an efficient classical description \cite{Gottesman1999, Aaronson2004}. Consequently, this motivates the study of quantum resources beyond entanglement. One such resource is magic, or nonstabilizerness \cite{Bravyi2005, Veitch2014} which characterizes quantum states lying outside the convex hull of stabilizer states \cite{Veitch2014, Howard2017}.\\

Magic plays a central role in providing a key resource for universal quantum computation \cite{Bravyi2005}. Preparation of stabilizer states, Clifford operations and Pauli measurements constitute resources that can be efficiently simulated on a classical computer. However, the injection of magic states promotes these easy operations to a universal set. More recently, magic has emerged as a complementary many-body diagnostic for probing quantum criticality, phase structure, and aspects of many-body states not captured by conventional observables \cite{Liu2020ManyBodyQM,White2021,Oliviero2022,Tarabunga2024,Falcao2025,Korbany2025,Zhang2026}.\\

Despite this progress, a central difficulty remains. Magic is generally challenging to characterize, both conceptually and computationally \cite{Heinrich2019,Hamaguchi2024}. Exact stabilizer decompositions and stabilizer-polytope optimizations become rapidly expensive as the number of qubits grows \cite{Aaronson2004,Heinrich2019,Hamaguchi2024}. This poses a particular challenge in many-body systems, where complete characterization of the full quantum state itself requires resources that grow exponentially with system size, motivating approaches based on restricted diagnostics such as local marginals or reduced density matrices (rdms) \cite{Cramer2010,Xin2017,Huang2020}.\\

In parallel, recent works have shown how magic can be detected from restricted sets of Pauli observables, using
projected stabilizer polytopes \cite{Varela2026} and, more recently, the
graph-theoretic structure of the accessible observables \cite{Liu2026}.
Together, these developments motivate several broader many-body questions, such as 
when magic becomes accessible through local information, what does it reveal about 
the underlying many-body physics?   
And can such restricted sets of observables ever offer a faithful representation
of the reduced density matrix under study?
\\

The Kitaev honeycomb model (KHM) provides a particularly natural setting in which to investigate 
how many-body information survives in local probes at finite temperature. 
This model offers a celebrated example of a quantum spin liquid (QSL) ground state, 
and is exactly solvable through a fractionalization of the spin degrees of freedom into 
Majorana fermions coupled to a static $\mathbb{Z}_2$ gauge field \cite{Kitaev2006}. 
At finite temperature, this fractionalized structure gives rise to two characteristic thermal crossover scales, associated respectively with itinerant Majorana fermions and thermal fluctuations of the $\mathbb{Z}_2$ fluxes \cite{Nasu2015, Yamaji2016, Yoshitake2016, Yoshitake2017_Majorana,Yoshitake2017_Temp, Hermanns2018, Rousochatzakis2019, Li2020, Feng2020, Gohlke2023}. 
A natural issue is then how much of this
emergent structure survives when the system is accessed only through small
subsystems. 
Recent work has shown that the thermal evolution of the KHM is
indeed reflected in its local entanglement structure, demonstrating that
small local marginals retain nontrivial information about the Kitaev spin-liquid both at zero \cite{Lyu2026} as well as finite temperature \cite{Sabharwal2025characterizing}. 
This raises the further question of whether quantum resources beyond entanglement encode complementary information about the same finite-temperature physics.\\

In this work, we address this question by studying magic in local marginals of the KHM at finite temperature. Using the robustness of magic (RoM) \cite{Howard2017}, we first determine when magic becomes locally accessible and then use the corresponding optimal witnesses to probe the structure underlying the local state. We find that these
witnesses are far from arbitrary. Instead, they organize into small, physically interpretable operator spaces built from the local correlations and symmetries of the Kitaev model, which close under the Jordan product to form finite-dimensional Euclidean Jordan algebras \cite{JordanVonNeumannWigner1934, FarautKoranyi1994}.
These operator spaces, in turn, define low-dimensional feature spaces that retain the local rdms with very high fidelity throughout their thermal evolution.\\

For the six-site hexagonal marginal, this connection can be made stronger: the relevant operator space forms the \emph{symmetry-resolved plaquette algebra} and, when the corresponding symmetries are realized exactly, provides an exact description of both the local state and its robustness of magic. Consequently, the same compact description also retains the information needed to characterize other local quantum resources, such as genuine multipartite entanglement (GME). The onset of local magic therefore does more than signal departure from the stabilizer polytope. It reveals a small set of physically meaningful operators that captures nontrivial aspects of the finite-temperature Kitaev spin liquid. Figure~\ref{fig:overview.schematic} summarizes this physical picture and the connections developed throughout the paper. This perspective is also experimentally timely, given recent progress toward digital simulations of the KHM using reconfigurable neutral-atom arrays \cite{Evered2025}, which provide a natural platform for probing these structures.\\

The remainder of this paper is organized as follows. We begin by
introducing the KHM in Sec.~\ref{subsec:KHM} and the stabilizer
framework in Sec.~\ref{subsec:Stabilizer.states.polytope.and.magic}, which we
use to present the robustness of magic in Sec.~\ref{subsec:RoM} and its witnesses in Sec.~\ref{subsec:Witness.as.hyperplane}. We then present our results in Sec.~\ref{sec:Results}. We first study the three-site marginal in Sec.~\ref{subsec:three.site.results} as an instructive example of the structure revealed by the optimal magic witnesses, before turning to our main case, the six-site hexagonal marginal, in Sec.~\ref{subsec:six.site.results}. There we show how the witness structure leads to a small symmetry-resolved operator space and establish its connection to the local state. 
We then show in Sec.~\ref{subsec:symmetry.exactness} how the six-site operator space follows from the symmetries of the model and establish the conditions under which the reduced robustness of magic is exactly equal to the full robustness of magic. We subsequently ask how much of the local thermal state is retained within these operator spaces in Sec.~\ref{subsec:thermal.state}, and then illustrate the broader applicability of the symmetry-resolved operator space by applying it to GME in Sec.~\ref{subsec:GME}. We close with a discussion in Sec.~\ref{sec:discussion} of the physical picture that emerges and its broader implications. Numerical details, stabilizer-polytope facet inequalities, the four-site witness analysis, and the Euclidean Jordan algebra structure of the six-site operator space are collected in the Appendices.

\section{Preliminaries}
In this section, we introduce the definitions and concepts used throughout the paper. 
We first recall the KHM studied in this work. 
We then review the stabilizer formalism, the notion of magic, and the robustness-of-magic witness construction. Finally, we explain how these witnesses will be interpreted as local Pauli observables.

\subsection{Kitaev honeycomb model}
\label{subsec:KHM}

\begin{figure}[t!]
    \centering
	\subfloat[ 24--site cluster \label{fig:honeycomb.finite.size}]{\includegraphics[height=0.4\linewidth]{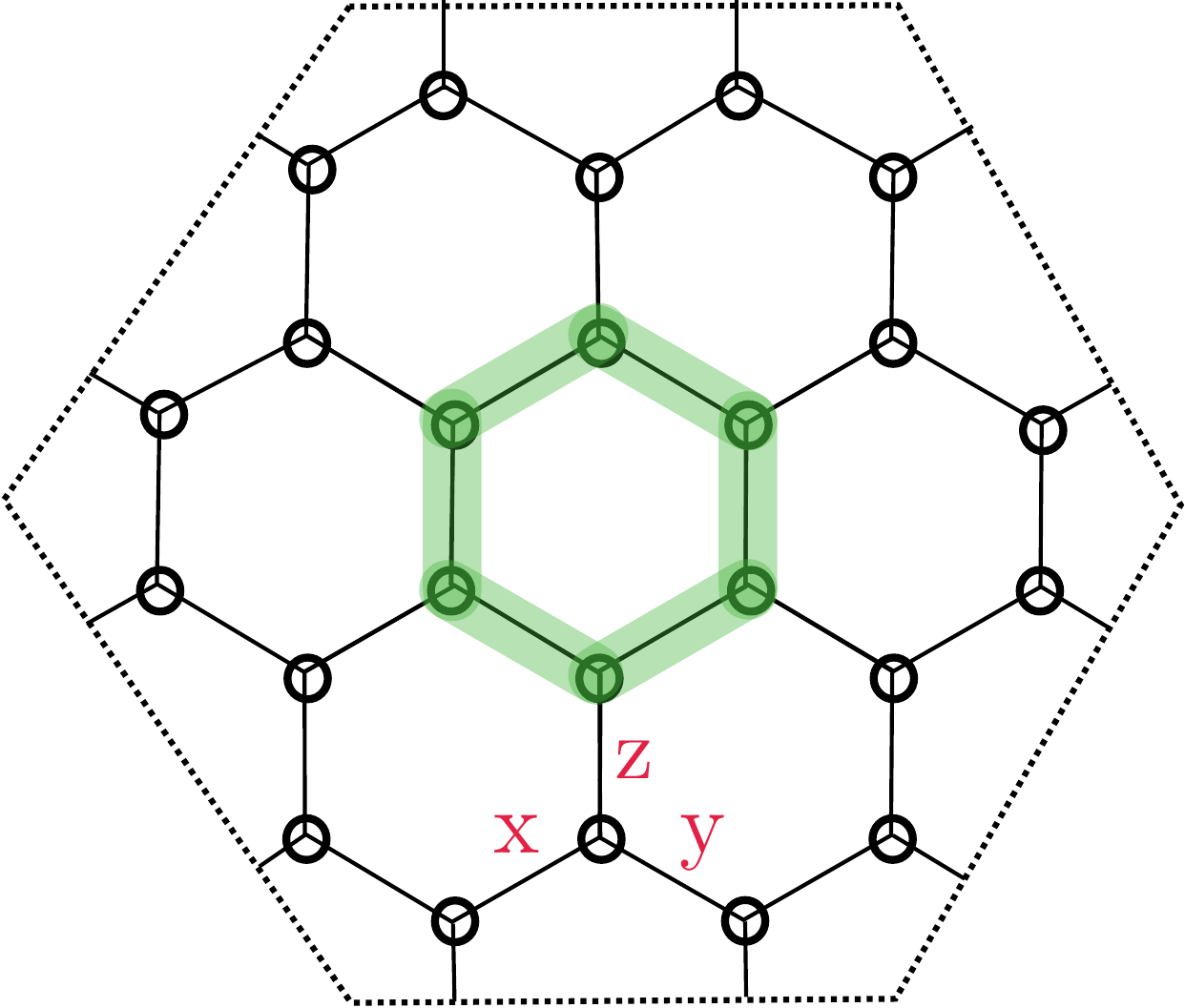}}
    \qquad
    \subfloat[ $n$-site residuals\label{fig:n.site.cluster}]
    {\includegraphics[height=0.4\linewidth]{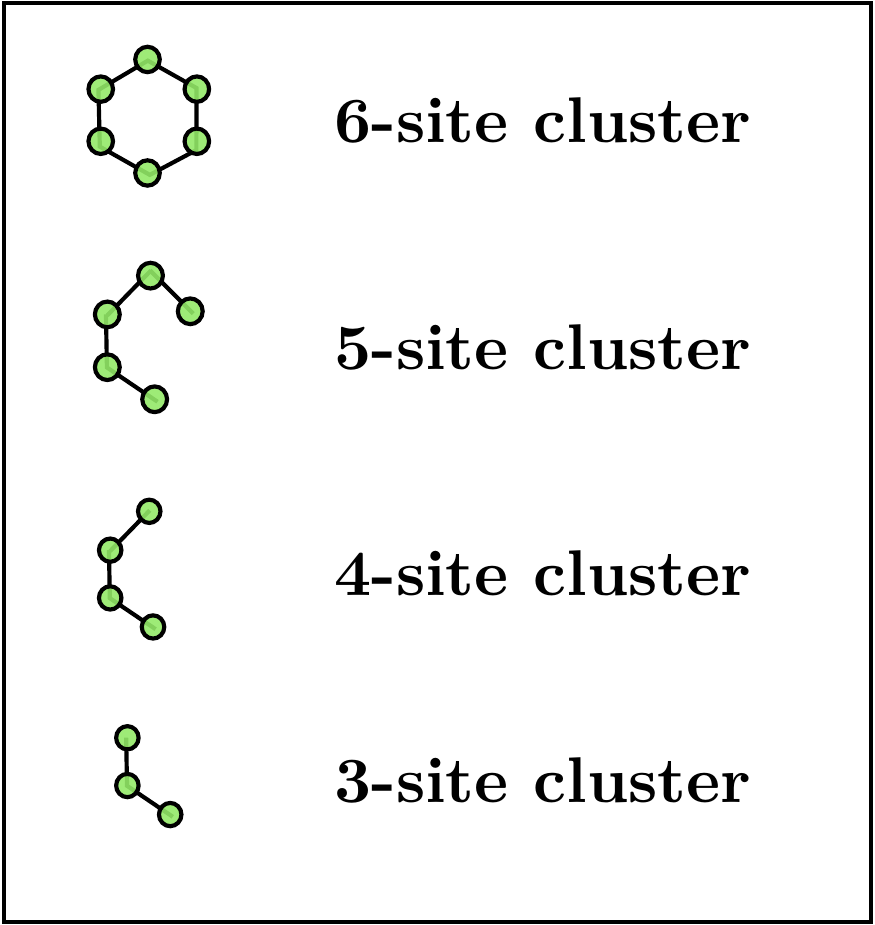}}
    \caption{
    Honeycomb lattice and sites associated with residuals used in this study. 
    (a) 24--site cluster used in simulations, possessing full $C_{6v}$ symmetry 
    of the honeycomb lattice.
    Feature spaces are defined on sites belonging 
    to the hexagonal cycle shown in green.
    Dashed lines denote the periodic boundary conditions of the cluster.
    (b) Clusters of spins within hexagonal cycle which are used in defining n--site residuals. 
    }
    \label{fig:cluster-for-KHM}
\end{figure}

%
We consider the spin-$1/2$ KHM under periodic boundary conditions
\begin{eqnarray}
\label{eq:H.Kitaev}
{\mathcal H}_{\sf KHM} 
	&=&    \frac{-1}{4}\sum_\alpha \sum_{\langle ij \rangle \in \{\alpha\}} J_\alpha \sigma_i^\alpha  \sigma_j^\alpha  \; , 
\end{eqnarray}
where the sum on $\alpha = \{x,y,x\}$ runs over the three inequivalent 
1$^{st}$--neighbour bonds of a honeycomb lattice, illustrated in Fig.~\ref{fig:cluster-for-KHM}, 
and we consider the isotropic parameter set 
\begin{eqnarray}
	  J_\alpha \equiv J > 0 \; . 
\end{eqnarray}
The exact solubility of the KHM model at $T=0$ rests on the existence 
of an extensive set of conserved quantites 
\begin{eqnarray}
\label{eq:W.plaquette}
P &=&  \sigma_1^y  \sigma_2^x  \sigma_3^z  \sigma_4^y  \sigma_5^x  \sigma_6^z  = \pm 1  \; , 
\end{eqnarray}
defined on each hexagonal plaquette of the honeycomb lattice.
In the ground state, these take values 
\begin{eqnarray}
P =  1  \qquad [T=0] \; .
\end{eqnarray}

At finite temperature, for a system in thermal equilibrium, the Gibb's state is
\begin{equation}
\rho(\beta)
=
\frac{e^{-\beta H}}{Z(\beta)},
\qquad
Z(\beta) =
\Tr\left(e^{-\beta H}\right)\;.
\label{eq}
\end{equation}
%
For a subsystem $A$, the local reduced density matrix (rdm) is
\begin{equation}
\rho_A(\beta)
=
\Tr_{\bar A}\rho(\beta),
\label{eq}
\end{equation}
where $\bar A$ denotes the complement of $A$. 

The central objects of this work are rdms, 
\begin{eqnarray}
	\rho^{(n)} \; , \quad n=\{3,4,5,6\} \; , 
\end{eqnarray}
defined on  clusters of $n$ spins, as illustrated in Fig.~\ref{fig:KHM}.
We will refer to these rdms interchangably as residuals.
These residuals, along with thermodynamic 
quantities such as heat capacity
 \begin{eqnarray}
 c(T) = \frac{1}{N} \frac{\langle \mathcal{H}_{\sf KHM}^2 \rangle 
		 - \langle \mathcal{H}_{\sf KHM} \rangle^2}{T^2} \; .
\label{eq:heat.capacity}.
 \end{eqnarray}
are are calculated using the method of thermal pure quantum 
states (TPQ)  \cite{Imada1986,Hams2000,Iitaka2003,Machida2012,Sugiura2013,Ikeuchi2015,Endo2018}, 
for a high--symmetry cluster of $N=24$ spins, 
as described in Appendix~\ref{Append:Numerics}.
\\

\subsection{Stabilizer states, polytope and magic}\label{subsec:Stabilizer.states.polytope.and.magic}
Having obtained the local rdms, we next ask whether they can be described within the stabilizer framework. More precisely, for each local marginal $\rho_A$, we ask whether it lies inside or outside the convex hull of stabilizer states. This convex set is known as the stabilizer polytope. 

A pure $n$-qubit stabilizer state $\ket{\phi}$ is defined as the simultaneous eigenstate of a set of $n$ independent and mutually commuting Pauli strings ${P_1,\ldots,P_m}$. These operators generate a state dependent stabilizer group $\mathcal{G} = \langle P_1, P_2,...,P_m\rangle$ and satisfy
\begin{equation}
P_j\ket{\phi}=\ket{\phi},
\qquad
j=1,\ldots,m.
\end{equation}
The state $\ket{\phi}$ is uniquely specified by its stabilizer group. Equivalently, its density operator can be written as
\begin{equation}
\zeta = \ket{\phi}\bra{\phi} =\frac{1}{2^n} \sum_{j=1}^n P_j.
\end{equation}

Let $\ms_{n}$ denote the set of pure stabilizer states $\zeta_{l}$ of $n$-qubits,
and $\mathrm{STAB}_n$ be the convex hull of such states,
\begin{align}
    \mathrm{STAB}_n &= \text{conv}\{\ms_{n}\} \nonumber \\
                    &=  \bigg\{\sum_{l}x_l\zeta_l \;\bigg|\; \zeta_{l} \in \mathcal{S}_{n}, x_l \geq 0, \sum_l x_l =1\bigg\}\;.
\end{align}
Thus, $\mathrm{STAB}_n$ is the stabilizer polytope, with the pure stabilizer states $\zeta_l$ forming its vertices. Note that the size of $\ms_n$ scales super-exponentially with $n$ \cite{Aaronson2004}
\begin{equation}
    |\ms_n| = 2^{n}\prod_{k=0}^{n-1} (2^{n-k} + 1) = 2^{O(n^2)}\;,
\end{equation}
thereby highlighting one of the main computational difficulties in characterizing magic for an increasing number of subsystems.

A state $\rho$ that lies inside the stabilizer polytope can be written as a probabilistic mixture of pure stabilizer states and is therefore said to be magic free or non-magical. Conversely,
\begin{equation}
\text{ If } \rho\notin\mathrm{STAB}_n \implies \rho \text{ possesses magic (non-stabilizer).} 
\end{equation}
\begin{figure}[t!]
    \centering
    \includegraphics[width=0.9\linewidth]{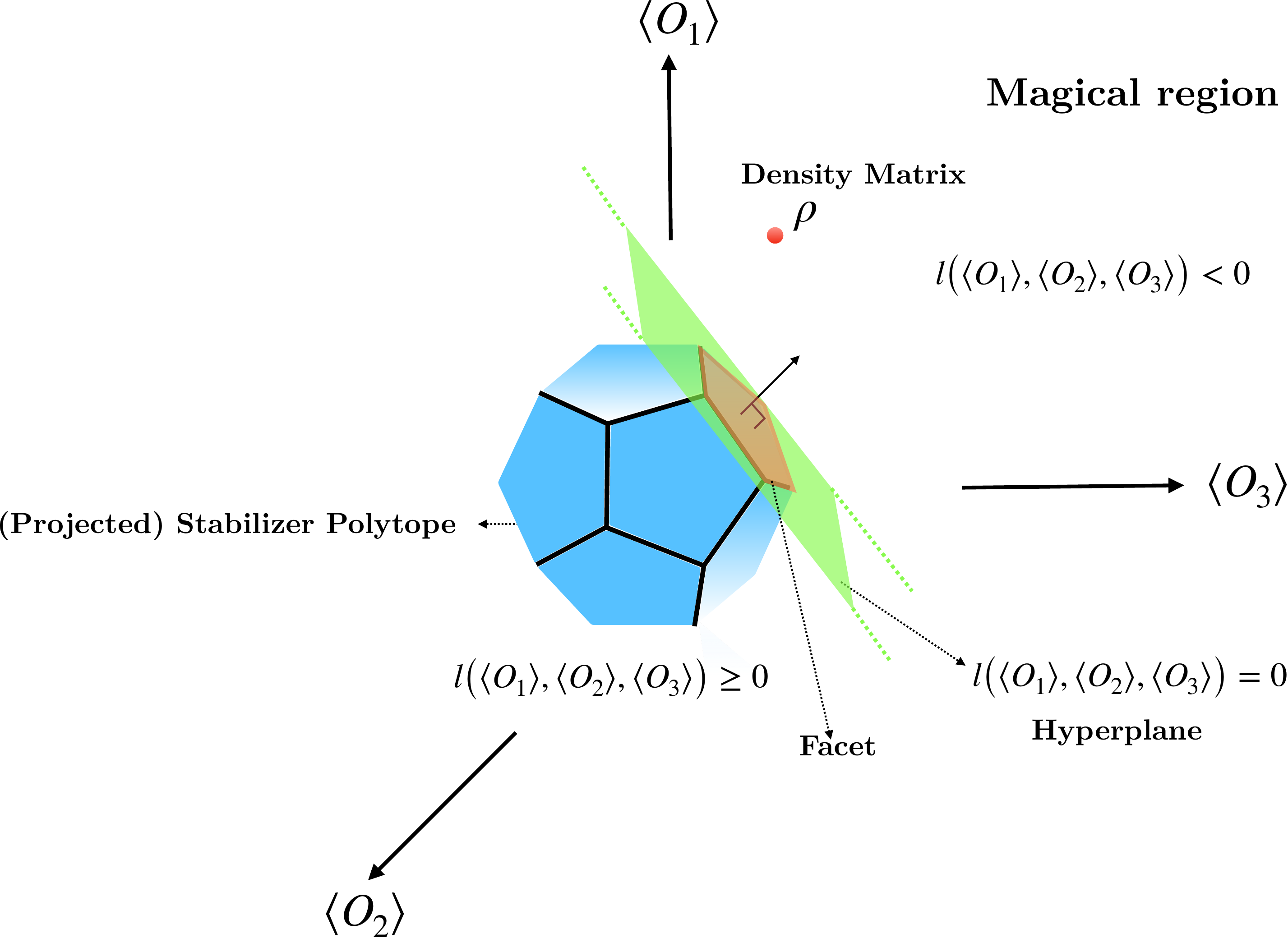}
    \caption{A schematic depiction of a three-dimensional feature space. The green plane represents a supporting hyperplane, while the orange face corresponds to its intersection with the projected stabilizer polytope, defining a facet of the polytope.}
    \label{fig:Stab.feature.space}
\end{figure}

\subsection{Robustness of magic: Primal and dual formulation}\label{subsec:RoM}
To quantify the amount of magic present in a state of interest $\rho$, we make use of robustness of magic (RoM) \cite{Howard2017,Hamaguchi2024},
\begin{equation}\label{eq:ROM}
    \mathcal{R}[\rho] = \min_{r}\bigg\{1+2r \;\bigg|\; \frac{\rho + r \tau}{1+r} \in \stab, \tau \in \stab, r \geq 0\bigg\}\;.
\end{equation}
Here, $r$ quantifies the amount of stabilizer noise that must be mixed with $\rho$ such that the resulting state lies within the stabilizer polytope.
RoM is a magic monotone and satisfies the following properties \cite{Howard2017},
\begin{itemize}
     \item Faithfulness: 
        \begin{equation}
            \mathcal{R}[\rho] = 1 \iff \rho \in \stab
        \end{equation}
    Otherwise, $\mathcal{R}[\rho] > 1$, which signals the presence of magic.
    \item Convexity:
    \begin{equation}\label{eq:conv.ROM}
        \mathcal{R}\big[\sum_{l}x_l \rho_l\big] \leq \sum_{l}x_l\mathcal{R}[\rho_l]
    \end{equation}
    \item Monotonicity:
    \begin{equation}
        \mathcal{R}[\mathcal{E}(\rho)]\leq \mathcal{R}[\rho]
    \end{equation}
    where, $\mathcal{E}$ are trace preserving stabilizer channels.
    \item Sub-multiplicativity:
    \begin{equation}
        \mathcal{R}[\rho_1 \otimes \rho_2] \leq \mathcal{R}[\rho_1].\mathcal{R}[\rho_2]
    \end{equation}
\end{itemize}

Eq.~\ref{eq:ROM} can equivalently be written as an $\ell_1$-norm minimization over pure stabilizer decompositions \cite{Howard2017,Hamaguchi2024},
\begin{align}\label{eq:ROM.l1}
\mathcal{R}[\rho] =\min_{\{x_l\}}
\bigg\{
\sum_l |x_l|
\; \bigg|\;
&\rho=
\sum_l x_l\zeta_l,
\zeta_l\in\ms_n,
x_l\in\mathbb{R},
\nonumber\\
&\sum_{l}x_l = 1\bigg \}\;,
\end{align}
which lends itself to the interpretation as a measure of the minimum quasiprobability negativity required to represent $\rho$ as a mixture of pure stabilizer states. The coefficients $x_l$ are not required to be positive and therefore define a stabilizer pseudomixture. For a stabilizer state, a decomposition with $x_l\geq0$ exists and $\mathcal{R}[\rho]=1$. For a magic state, negative coefficients are necessary and $\mathcal{R}[\rho]>1$. For practical computations, the $\ell_1$-norm optimization in Eq.~(\ref{eq:ROM.l1}) can be reformulated as a standard linear program (LP),
\begin{equation}\label{eq:ROM.stand.primal.LP}
\mathcal{R}[\rho] = \min_{\mathbf{z}} \{\mathbf{1}^T\mathbf{z} \;\big|\; A\mathbf{x} = \mathbf{b}, -\mathbf{z} \leq \mathbf{x} \leq \mathbf{z} \}\;,
\end{equation}
where 
\begin{equation}
b_j=\Tr[P_j\rho],
\qquad
A_{jl}=\Tr[P_j\zeta_l],
\end{equation}
are the expansions of the target and pure stabilizer states in the Pauli basis $\{P_j \}$. Eq.~\ref{eq:ROM.stand.primal.LP} can be solved using standard LP solvers like Gurobi \cite{Gurobi2025}. Furthermore, due to the fact the above optimization is feasible and bounded, strong duality provides an equivalent dual formulation of the problem,
\begin{equation}
    \mathcal{R}[\rho] = \max_{\mathbf{y}} \{\mathbf{b}^T\mathbf{y}\; \big|\; -\mathbf{1} \leq A^T \mathbf{y} \leq \mathbf{1}\}\;.
\end{equation}
Associating the vector $\mathbf{y}$ with the Hermitian operator
\begin{equation}
W
=
\sum_j y_jP_j,
\end{equation}
the dual constraints imply, 
\begin{equation}
\big|\Tr[W\zeta_l]\big| \leq 1, \qquad \forall \;\zeta_l\in\ms_n.
\end{equation}
At the optimum, we have
\begin{equation}
    \Tr[W^{*}\rho] = \mathcal{R}[\rho]\;.
\end{equation}
Importantly, $W^*$ is the optimal magic witness and it is this structure of this operator that forms the central focus in the present work. We make use of the efficient implementation as described in \cite{Hamaguchi2024}. In the following, without any loss of generality, we simply use $W$ to mean the optimal witness as opposed to $W^*$. 

\subsection{Witness as a supporting hyperplane}\label{subsec:Witness.as.hyperplane}
The witness $W$, together with the stabilizer bound, defines the hyperplane
\begin{equation}
    \Tr[W\omega] = 1\;.
\end{equation}
with $\omega$ being an arbitrary $n$-qubit state. Equivalently, 
by defining a linear witness function
\begin{equation}\label{eq:linear.Witness.function}
    \ell[\omega] = 1- \Tr[W\omega]\;,
\end{equation}
the equation
\begin{equation}
    \ell[\omega]=0\;.
\end{equation}
then describes the same supporting hyperplane, as shown schematically in Fig. \ref{fig:Stab.feature.space}.
The dual constraints ensure that
\begin{equation}
\ell[\sigma]\geq0,
\qquad
\forall \; \sigma\in\stab,
\end{equation}
whereas 
\begin{equation}
    \ell[\rho]<0
\end{equation}
certifies that the target state $\rho$ lies outside the stabilizer polytope and therefore possesses magic, as in Fig.~\ref{fig:Stab.feature.space}. The intersection of the supporting hyperplane with the stabilizer polytope defines an exposed face.

\section{Results: Emergence of structured magic}
\label{sec:Results}


\begin{figure}[t!]

\counterwithin*{subfigure}{figure}   
\renewcommand{\thesubfigure}{(\alph{subfigure})}
\makeatletter
\renewcommand{\p@subfigure}{\the\numexpr\value{figure}+1\relax}
\makeatother

\newcommand{\panel}[2]{
  \refstepcounter{subfigure}\label{#1}%
  \makebox[1.6em][l]{(\alph{subfigure})}%
  \adjustbox{valign=t}{\includegraphics[width=0.8\columnwidth]{#2}}\par\vspace{2pt}}

  \centering
  
  \panel{fig:phase.diagram}{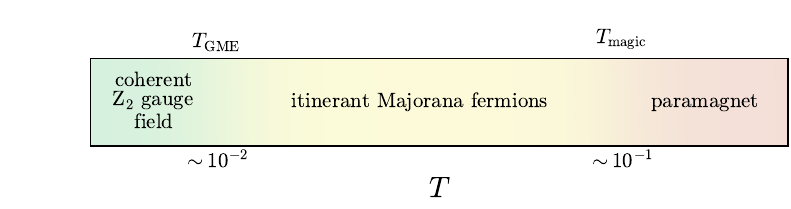}
  \panel{fig:magic}{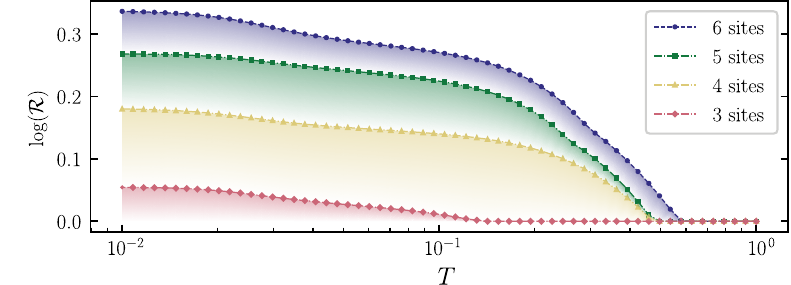}
  \panel{fig:GME}{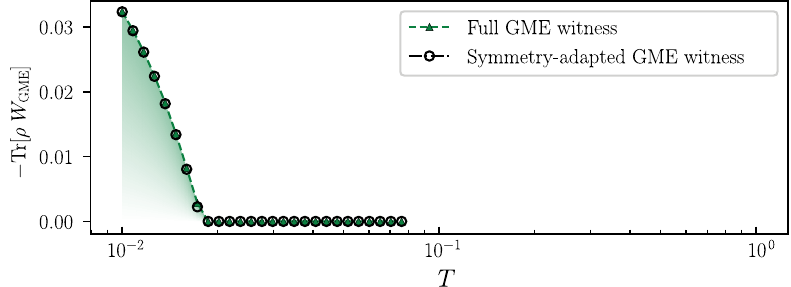}
  \panel{fig:fidelity}{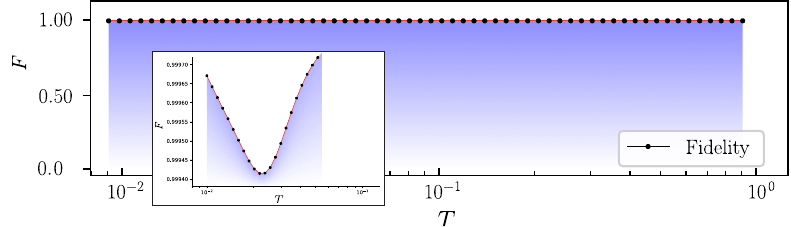}
  \panel{fig:plaquette}{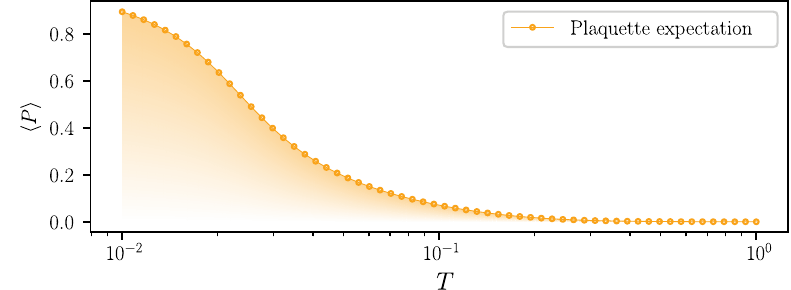}
  \panel{fig:heat.capacity}{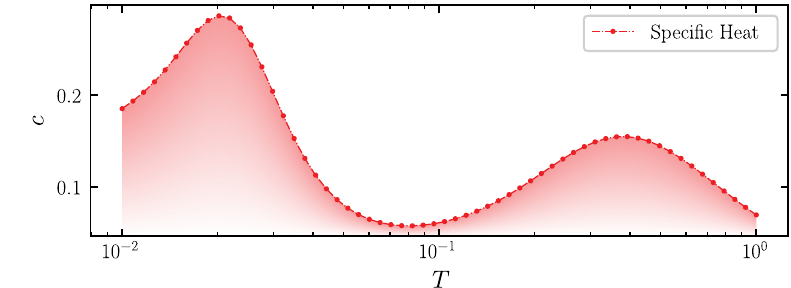} 
  
  \caption{Evolution of magic and related physical quantities with temperature 
  in the Kitaev honeycomb model (KHM).
  (a)~Schematic phase diagram, following \cite{Nasu2015,Sabharwal2025characterizing}, showing 
  how the fractionalized spin excitations of the KHM evolve with temperature.  
  (b)~Logarithmic robustness of magic (RoM) for the marginals $\rho^{(n)}$
  with $n=3,4,5,6$  [Fig.~\ref{fig:cluster-for-KHM}], 
  showing onset of magic at $T_{\sf magic} \gtrsim 0.5$. 
  (c)~Equivalent results for genuine multipartite entanglement (GME) within the six--site marginal, $\rho^{(6)}$, 
  as discussed in Section~\ref{subsec:GME}, 
  showing onset at $T_{\sf GME} \approx 0.02$.  
  Results are shown for both the full GME witness, and the witness evaluated within the reduced feature space 
  (symmetry--adapted witness).  
  (d)~Fidelity of 6--site marginal evaluated within reduced feature space, 
  as discussed in Section~\ref{subsec:thermal.state}.   
  %
  Deviations from $F=1$ are maximally of order $10^{-3}$ [Inset].    
  (e)~Expectation value of plaquette operator, $\langle P\rangle$ [Eq.~(\ref{eq:W.plaquette})], 
  showing $\langle P \rangle \to 1$ for $T \to 0$ as expected for the quantum spin liquid (QSL) ground state.
  (f)~Heat capacity $c(T)$ [Eq.~(\ref{eq:heat.capacity})], showing peaks at characteristic temperature scales.  
  All results are taken from thermal pure quantum state (TPQ) calculations for the Kitaev honeycomb 
  (KHM) model, Eq.~(\ref{eq:H.Kitaev}), defined on a symmetric cluster of $N=24$ spins, as described in the text.}
  
  \label{fig:rom.temperature}
  
\end{figure}



We now investigate the magic contained in the local marginals of the KHM at finite temperature. Here we focus primarily on the three, four, and six-site marginals, obtained using the TPQ method, Appendix.~\ref{Append:Numerics}, for which we analyze the structure of the witnesses near the temperature at which magic emerges and subsequently examine their behavior with temperature.\\

In Fig.~\ref{fig:rom.temperature} we show results for the logarithmic RoM,  
$\log (\mathcal{R}[\rho^{(n)}])$, for marginals $\rho^{(n)}$ with $n=3,4,5,6$.
For later comparison we also provide a schematic phase diagram [Fig.~\ref{fig:phase.diagram}],  together with results for genuine multipartite entanglement (GME)  [Fig.~\ref{fig:GME}]  fidelity  [Fig.~\ref{fig:fidelity}], plaquette operator $W$  [Fig.~\ref{fig:plaquette}],  and specific heat $c(T)$ [Fig.~\ref{fig:heat.capacity}].

At high temperature, all four marginals are stabilizer
mixtures and therefore have vanishing magic [Fig.~\ref{fig:magic}]. 
As the system is cooled, magic
first becomes detectable in the larger marginals and subsequently appears in
smaller clusters.
In what follows, we concentrate on the instructive cases of three-- and six--site marginals, 
with further details of four-site marginals given in Appendix~\ref{Append:Half.plane.four.site}.\\

Having established when magic becomes locally accessible, we next ask what
physical correlations are responsible for it.  We find that the witnesses are not arbitrary combinations of Pauli operators. Instead, their Pauli terms combine into compact algebraic structures involving commuting sector operators and non-commuting dressed operators. For the six-site hexagon, this organization becomes more explicit. Here the witness terms are organized into symmetry-related combinations of bond, loop, and string operators.\\

The repeated organization of the Pauli terms in the witness into compact composite operators motivates us to identify an operator space that captures the structure of the witnesses. Rather than working with the complete Pauli expansion of each local marginal, we group the Pauli strings appearing in the witnesses into a small set of composite Hermitian operators $O_a$, which take the form
\begin{equation}\label{eq:composite.operator}    
O_a=\sum_{j} q_{ja}P_{j} \;, a \in \{1,..n_f\}
\end{equation}
where $P_{j}$ denotes a $n$-qubit Pauli string and $q_{ja}\in\mathbb{R}$. We denote the real linear span of these operators by
\begin{equation}
\mathcal{V}= \text{span}_{\mathbb{R}}\{I,O_1,\ldots,O_{n_{f}}\}.
\end{equation}
This construction is closely related to the restricted-measurement framework of Varela \textit{et al.}~\cite{Varela2026}, where a set of accessible Pauli measurements $\mathcal{M}$ defines a projection of the stabilizer polytope onto the corresponding expectation values. Here, we use $\mathcal{V}$ to denote the operator space generated by the chosen `features'.\\

The expectation values of these operators
\begin{equation}
\mathbf{x}(\omega)
=
\left(
\langle O_1\rangle_{\omega},
\ldots,
\langle O_{n_{f}}\rangle_{\omega}
\right),
\end{equation} then define a corresponding `feature space', where
\begin{equation}
\langle O_j\rangle_{\omega}=\Tr[O_j\omega]\;.
\end{equation}
This construction substantially reduces the number of observables required to describe the local magic and allows us to identify which correlations drive the departure of the marginal from the stabilizer polytope.\\

Moreover we find that beyond providing a compact representation of the witnesses, the resulting operator spaces possess a nontrivial algebraic structure. We show that the operator spaces obtained for the three, four, and six-site marginals are closed under the Jordan product (i.e. the anti-commutator)
\begin{equation}\label{eq:JordanProduct} 
A\circ B
=
\frac{1}{2}
\left(
AB+BA
\right).
\end{equation}
Thus,
\begin{equation}
A,B\in\mathcal{V}
\quad\Longrightarrow\quad
A\circ B\in\mathcal{V}.
\end{equation}
Equipped with the Hilbert--Schmidt (HS) inner product
\begin{equation}\label{eq:HSproduct}
\langle A,B\rangle
=
\Tr[AB]\;,
\end{equation}
each $\mathcal{V}$ therefore forms a finite-dimensional Euclidean Jordan algebra (EJA) \cite{JordanVonNeumannWigner1934,FarautKoranyi1994}. The composite operators identified from the witnesses, consequently, do not constitute merely an ad hoc collection of Pauli terms but generate closed algebraic structures naturally adapted to the corresponding local marginals.\\

We first illustrate this construction for the three-site marginals, where the witnesses and their associated operator algebras admit particularly compact decompositions. We then turn to the six-site hexagon, where the operators $O_{a}$ are further organized into symmetry-related combinations generated by the local Kitaev bond algebra. We subsequently quantify how accurately the projection onto $\mathcal{V}$ reconstructs the physical reduced density matrices. Finally, by evaluating the corresponding feature vectors for all pure stabilizer states, we construct the projected stabilizer polytopes and analyze the facets crossed by the finite-temperature trajectories.

\subsection{Pedagogical examples: Witness and algebraic structure of the three and four-site marginals}\label{subsec:three.site.results}
\begin{figure*}[t!]
    \centering
    \subfloat[Three-site trajectory\label{fig:3site-trajectory}]{
        \begin{minipage}[t][0.36\textwidth][t]{0.48\textwidth}
            \vspace{0pt}
            \includegraphics[width=0.9\linewidth]{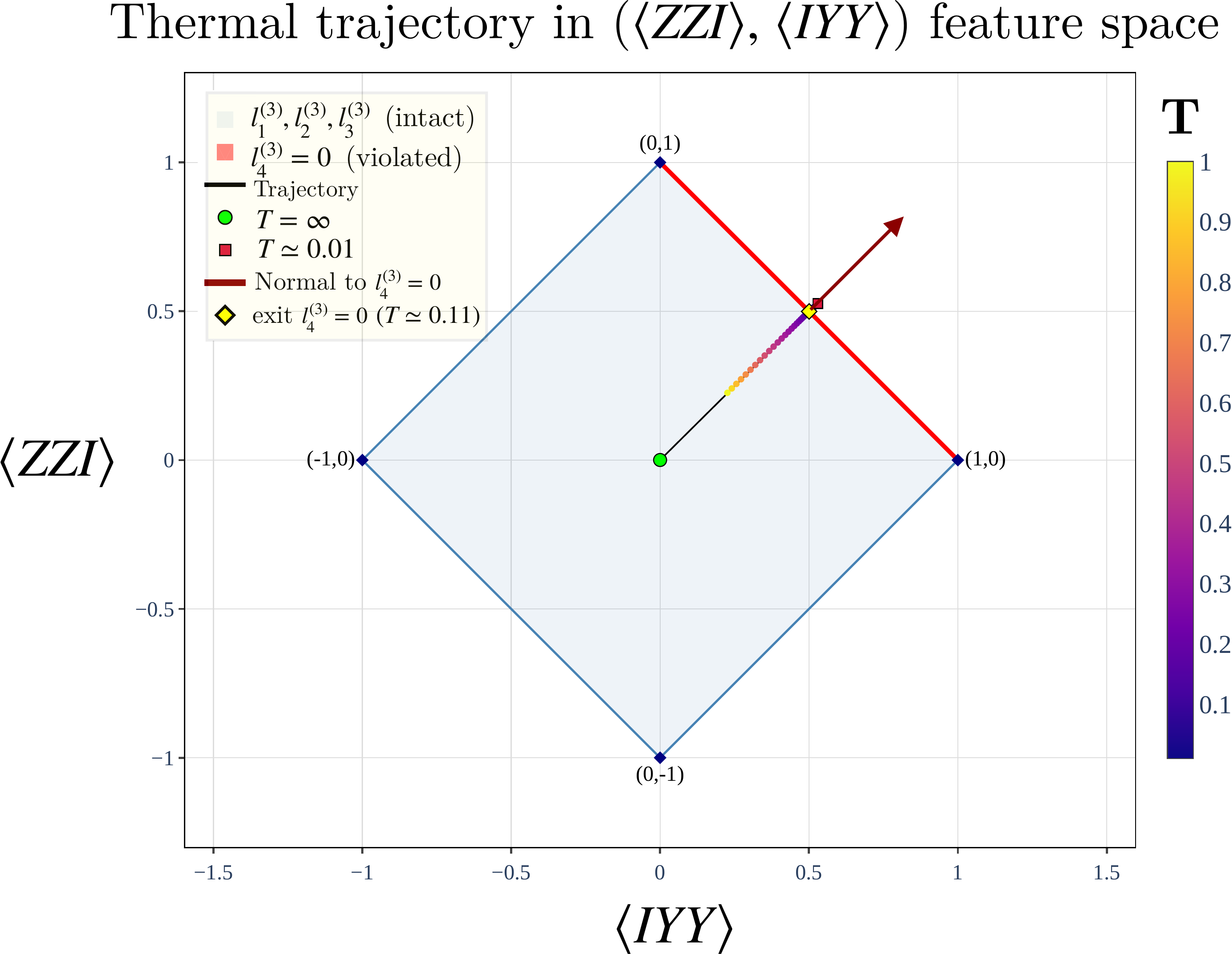}
        \end{minipage}
    }
    \subfloat[Four-site trajectory\label{fig:4site-trajectory}]{
        \begin{minipage}[t][0.36\textwidth][t]{0.48\textwidth}
            \centering
            \vspace{0pt}
            \includegraphics[width=\linewidth]{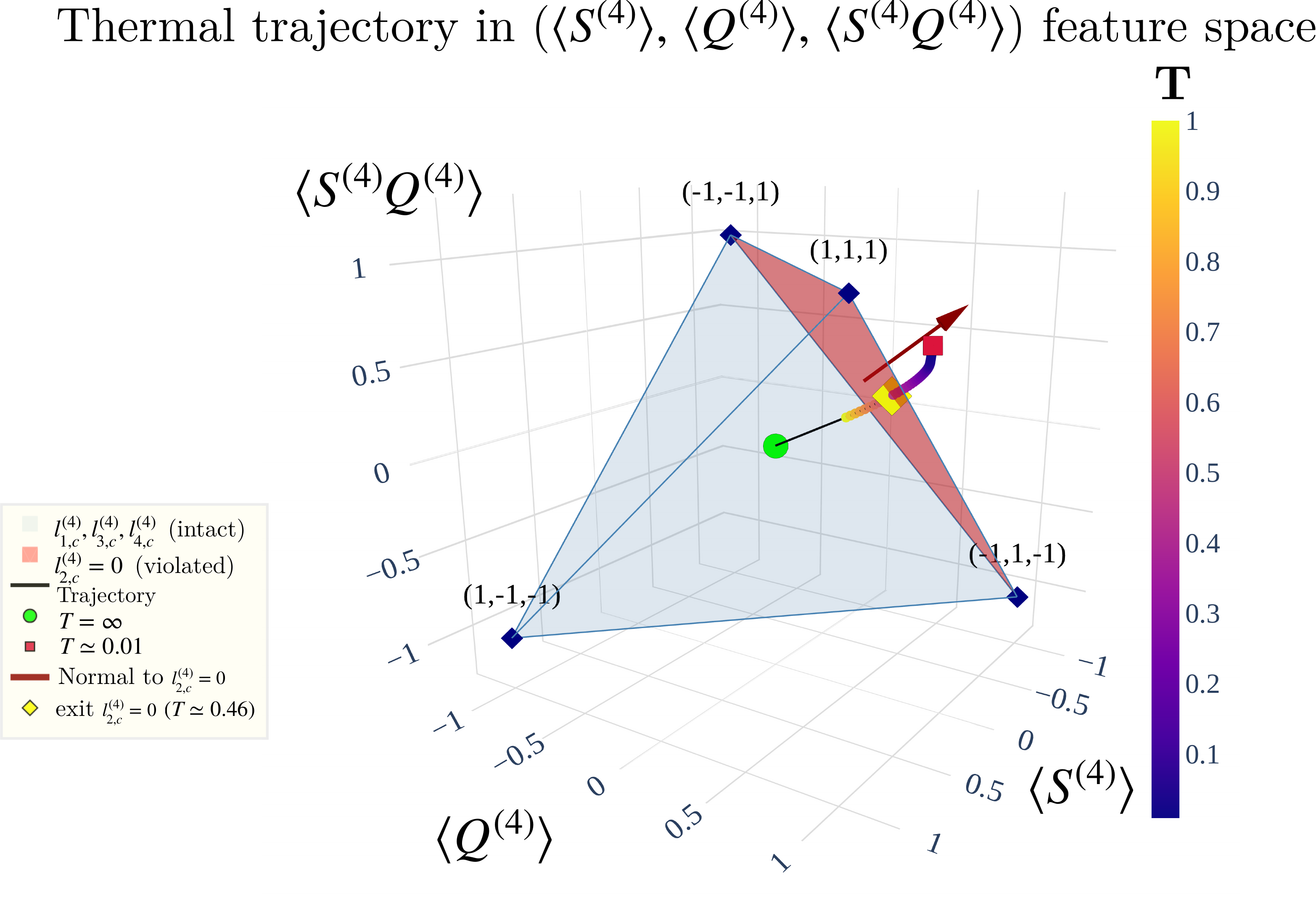}
        \end{minipage}
    }
\caption{Thermal trajectories of the three and four-site reduced density matrices in their corresponding feature spaces. In both cases, the $T=\infty$ (green circle) state lies at the center of the stabilizer polytope, a diamond (three-site) and projected stabilizer polytope, a tetrahedron (four-site).  Upon cooling, the thermal trajectory moves toward the boundary and eventually crosses the highlighted red facet, marking the onset of magic. (a) Three-site marginal in the feature space $(\langle ZZI\rangle,\langle IYY \rangle)$. The yellow diamond marks the trajectory at $T\simeq0.11,$, where it reaches the facet supported by the hyperplane in Eq.~\eqref{eq:three.site.hyperplane}.  (b) Four-site marginal in the feature space $(\langle S^{(4)}\rangle,\langle Q^{(4)}\rangle,\langle S^{(4)}Q^{(4)}\rangle)$. The yellow diamond marks the corresponding boundary point at $T\simeq0.46$, associated with the hyperplane in Eq.~\eqref{eq:four.site.hyperplane}.}
\label{fig:site-trajectories}
\end{figure*}

For the three-site marginal, the witness at the onset temperature turns out to be quite simple,
\begin{subequations}
\begin{align}
    W^{(3)} &= \frac{1}{2}(III + Q^{(3)}),\;\\
    Q^{(3)} &= A_1 + A_2,\;\\
    A_1 &= ZZI,\; A_2 = IYY 
\end{align}
\end{subequations}
depending only on the two bond operators $A_1$ and $A_2$, satisfying,
\begin{subequations}\label{eq:Gamma.three.site}
\begin{align}
    \{A_1,A_2\} &= 0 \implies A_1 \circ A_2 = 0\;\\
    A_i^2 &= I \implies (Q^{(3)})^2 = 2I\;.
\end{align}
\end{subequations}
These algebraic relations motivate the operator space
\begin{equation}
    \mathcal{V}^{(3)} = \text{span}_{\mathbb{R}}\{I, A_1, A_2 \}\;,
\end{equation}
wherein all the operators of this space are Hermitian. The operator space $\mathcal{V}^{(3)}$ is not closed under ordinary matrix multiplication. For example, $A_1A_2 = -iZXY$, does not lie within $\mathcal{V}^{(3)}$. However, it is closed under the Jordan product Eq.~\ref{eq:JordanProduct}. 
Since every normalized state satisfies 
\begin{equation}
\langle I\rangle_{\omega}=\Tr[\omega]=1\;,
\end{equation}
the identity component is fixed and does not provide an independent feature. The remaining two operator expectation values 
\begin{equation}
   \mathbf{x}^{(3)}(\omega) = \bigg(\langle A_1 \rangle_{\omega}, \langle A_2 \rangle_{\omega}\bigg)
\end{equation}
therefore, define the feature space of the three-site marginal.
After identifying the feature space, we evaluate it for every pure three-qubit stabilizer state $\zeta_l$. Each such feature vector $\mathbf{x}^{(3)}(\zeta_l)$ corresponds to a point in the two-dimensional feature space. Their convex hull defines a (projected) \footnote{For the three site marginal it is the full stabilizer polytope and not it's projection.} stabilizer polytope
\begin{equation}
    \mathcal{P}^{(3)} = \text{conv}\{\mathbf{x}^{(3)}(\zeta_l)\; | \; \zeta_l \in \mathcal{S}_{3})\} \subset \mathbb{R}^2\;.
\end{equation}
However, to determine where the marginal lies relative to this polytope, it is more convenient to describe the polytope as an intersection of half-spaces \cite{Ziegler1993}. We therefore introduce facet functions
\begin{equation}
    l_{\mu}^{(3)}(\mathbf{x}) = b_{\mu} + \mathbf{a}^{T}_{\mu}\mathbf{x}^{(3)} 
\end{equation}
such that the corresponding facet inequalities 
\begin{equation}
    l_{\mu}^{(3)}(\mathbf{x}) \geq 0 
\end{equation}
define a closed half-space, and whose boundary
\begin{equation}
      l_{\mu}^{(3)}(\mathbf{x}) = 0 
\end{equation}
defines a hyperplane
\begin{equation}
    \mathcal{H}_{\mu} =\{\mathbf{x} \in \mathbb{R}^{2} \;|\; l_{\mu}^{(3)}(\mathbf{x}) = 0\}\;.
\end{equation}
The polytope can thus be described as an intersection of these half-spaces
\begin{equation}
    \mathcal{P}^{(3)} = \bigcap_{\mu}\{\mathbf{x}^{(3)} \in \mathbb{R}^2\;|\; l_{\mu}^{(3)}(\mathbf{x}) \geq 0\}\;.
\end{equation}
The $\mu$-th boundary facet is the intersection of the polytope with the corresponding supporting hyperplane
\begin{equation}
    \mathcal{F}^{(3)}_{\mu} = \mathcal{P}^{(3)}\cap\mathcal{H}_{\mu}
\end{equation}
 Using \texttt{pycddlib} \cite{Troffaes_pycddlib_2025}, we convert the vertex representation obtained from the pure stabilizer feature vectors into the corresponding half-space representation.  
Moving forward, for an arbitrary state $\omega$, we introduce the following shorthand
\begin{equation}
    l^{(n)}_{\mu}[\omega] \equiv l^{(n)}_{\mu}(\mathbf{x}(\omega))\;,\quad n = 3,4,6\;.
\end{equation}
Every stabilizer state $\sigma$ satisfies
\begin{equation}
    l_{\mu}^{(3)}[\sigma] \geq 0 \;\;\forall\; \mu
\end{equation}
Therefore, if the three-site marginal violates at least one of these inequalities,
\begin{equation}
    l^{(3)}_{\mu}[\rho^{(3)}] < 0
\end{equation}
for some $\mu$, then it lies outside the polytope, thereby certifying the existence of magic. 
For the three-site marginal, we find the polytope to be a diamond [Fig.~\ref{fig:3site-trajectory}] specified according to the following four facet-inequalities, 
\begin{subequations}
    \begin{align}
    l_{1}^{(3)} &= 1+ \langle A_1 \rangle_{\omega} - \langle A_2 \rangle_{\omega} \geq 0\\
    l_{2}^{(3)} &= 1+ \langle A_1 \rangle_{\omega} + \langle A_2 \rangle_{\omega} \geq 0\\
    l_{3}^{(3)} &= 1- \langle A_1 \rangle_{\omega} + \langle A_2 \rangle_{\omega} \geq 0 \\
    l_{4}^{(3)} &= 1- \langle A_1 \rangle_{\omega} - \langle A_2 \rangle_{\omega} \geq 0\;.
    \end{align}
\end{subequations}
Of these, only one is violated and that one corresponds to the magic witness
\begin{equation}
    \ell[\omega] = \frac{l^{(3)}_{4}[\omega]}{2}
\end{equation}
The onset of magic corresponds to the thermal trajectory reaching the boundary facet contained in the supporting hyperplane
\begin{equation}\label{eq:three.site.hyperplane}
    l_{4}^{(3)} = 0\;,
\end{equation}
and subsequently, leaving the polytope as shown in Fig.~\ref{fig:3site-trajectory}.\\
 
For the four-site marginal $\rho^{(4)}$, we follow the same approach as for the three-site marginal.
The main difference is the additional central operator, which commutes with the remaining operators and in turn resolves the operator space into two sectors. We find the witness at the onset temperature to be given by
\begin{subequations}
\begin{align}
    W^{(4)}&=\Pi_{-}^{(S)}+
\Pi_{+}^{(S)}Q^{(4)},
\label{eq:witness.four.site}\\
\Pi_{\pm}^{(S)}&=\frac{I\pm S^{(4)}}{2},\\
S^{(4)}&=XXYY,\\
Q^{(4)}&=\Gamma_1^{(4)}+\Gamma_2^{(4)},\label{eq:Gamma.four.site}
\end{align}
\end{subequations}
where 
\begin{align}
&\Gamma_1^{(4)}=XXII,\;
\Gamma_2^{(4)}=XYXY
\end{align}
The operator $S^{(4)}$ commutes with each of the $\Gamma_a$
\begin{equation}
    [S^{(4)},\Gamma^{(4)}_a] = 0, \qquad a = 1,2
\end{equation}
while the $\Gamma$ operators anticommute
\begin{equation}
    \{\Gamma_{a}^{(4)}, \Gamma_{b}^{(4)}\} = 2\delta_{ab}I
\end{equation}
With the above algebraic relations, we define the operator space to be
\begin{align}
    \mathcal{V}^{(4)} = \text{span}_{\mathbb{R}}\{I,S^{(4)}, \Gamma_{1}^{(4)},\Gamma_{2}^{(4)},S^{(4)}\Gamma_{1}^{(4)},  S^{(4)}\Gamma_{2}^{(4)}\}\;,
\end{align}
which is closed under the Jordan product Eq.~\ref{eq:JordanProduct}. The nontrivial Jordan products between the basis operators are
\begin{subequations}
\begin{align}\label{eq:Jordan.Algebra.four.site}
S^{(4)}\circ S^{(4)}&=I,\\
S^{(4)}\circ\Gamma_a^{(4)}&=
S^{(4)}\Gamma_a^{(4)},\\
S^{(4)}\circ
(S^{(4)}\Gamma_a^{(4)})&=\Gamma_a^{(4)},\\
\Gamma_a^{(4)}\circ\Gamma_b^{(4)}&=\delta_{ab}I,\\
(S^{(4)}\Gamma_a^{(4)})\circ (S^{(4)}\Gamma_b^{(4)})&=\delta_{ab}I,\\
\Gamma_a^{(4)}\circ (S^{(4)}\Gamma_b^{(4)})&=\delta_{ab}S^{(4)}.
\end{align}
\end{subequations}
Equipped with the HS inner product Eq.~\ref{eq:HSproduct}, the operator space, $\mathcal{V}^{(4)}$ forms a six-dimensional Euclidean Jordan algebra (EJA). The algebra can be decomposed into the two eigensectors of $S^{(4)}$. Defining
\begin{equation}
    \mathcal{V}^{(4)}_{\pm} =\text{span}_{\mathbb R} \{\Pi_{\pm}^{(S)},\Pi_{\pm}^{(S)}\Gamma_1^{(4)}, \Pi_{\pm}^{(S)}\Gamma_2^{(4)}\}
\end{equation}
we obtain the direct sum decomposition
\begin{equation}
    \mathcal{V}^{(4)} = \mathcal{V}^{(4)}_{+} \oplus \mathcal{V}^{(4)}_{-}\;.
\end{equation}
Following the same procedure as the three-site marginal we can obtain the facet inequalities. The full set of $8$ inequalities are listed in Table~\ref{tab:four.site.facets}, all of which are found to be non-trivial. Furthermore, we can look at the compressed feature space. By noting that at the onset temperature, the witness depends only on
\begin{equation}\label{eq:four.site.feature.axes}
    \langle S^{(4)}  \rangle, \qquad  \langle Q^{(4)}  \rangle, \qquad   \langle S^{(4)} Q^{(4)} \rangle\;,
\end{equation}
we can define a compressed feature space vector:
\begin{equation}
    x_{c}^{(4)}(\omega) = (\langle S^{(4)}  \rangle_{\omega}, \langle Q^{(4)}  \rangle_{\omega}, \langle S^{(4)} Q^{(4)} \rangle_{\omega})\;,
\end{equation}
wherein the resulting polytope is a tetrahedron, defined by the four facet inequalities:
\begin{subequations}
\begin{align}
    l_{1,c}^{(4)}[\omega] &= 1+\langle S^{(4)} \rangle_{\omega} + \langle Q^{(4)}\rangle_{\omega} +\langle S^{(4)} Q^{(4)} \rangle_{\omega} \geq 0, \\
    l_{2,c}^{(4)}[\omega] &= 1+\langle S^{(4)} \rangle_{\omega} - \langle Q^{(4)}\rangle_{\omega} -\langle S^{(4)} Q^{(4)} \rangle_{\omega} \geq 0,  \\
    l_{3,c}^{(4)}[\omega] &= 1-\langle S^{(4)} \rangle_{\omega} + \langle Q^{(4)}\rangle_{\omega} -\langle S^{(4)} Q^{(4)} \rangle_{\omega} \geq 0,\\
    l_{4,c}^{(4)}[\omega] &= 1-\langle S^{(4)} \rangle_{\omega} - \langle Q^{(4)}\rangle_{\omega} +\langle S^{(4)} Q^{(4)} \rangle_{\omega} \geq 0 \;.
\end{align}
\end{subequations}
Moreover, of these, only one is violated and that corresponds to the magic witness
\begin{equation}
    \ell[\omega] = \frac{l^{(4)}_{2,c}[\omega]}{2}
\end{equation}

As shown in Fig.~\ref{fig:4site-trajectory}, the onset of magic corresponds to the thermal trajectory reaching the boundary facet contained in the supporting hyperplane
\begin{equation}\label{eq:four.site.hyperplane}
    l_{2,c}^{(4)} = 0\;,
\end{equation}
and subsequently, leaving the polytope. 

\subsection{Witness structure of the six site marginal}\label{subsec:six.site.results}
\begin{figure*}[t!]
    \centering
    \subfloat[Facet $l_{-}^{(6)}$: $P=-1$ sector
    \label{fig:6site-trajectory.facet91}]{
        \includegraphics[width=0.48\textwidth]{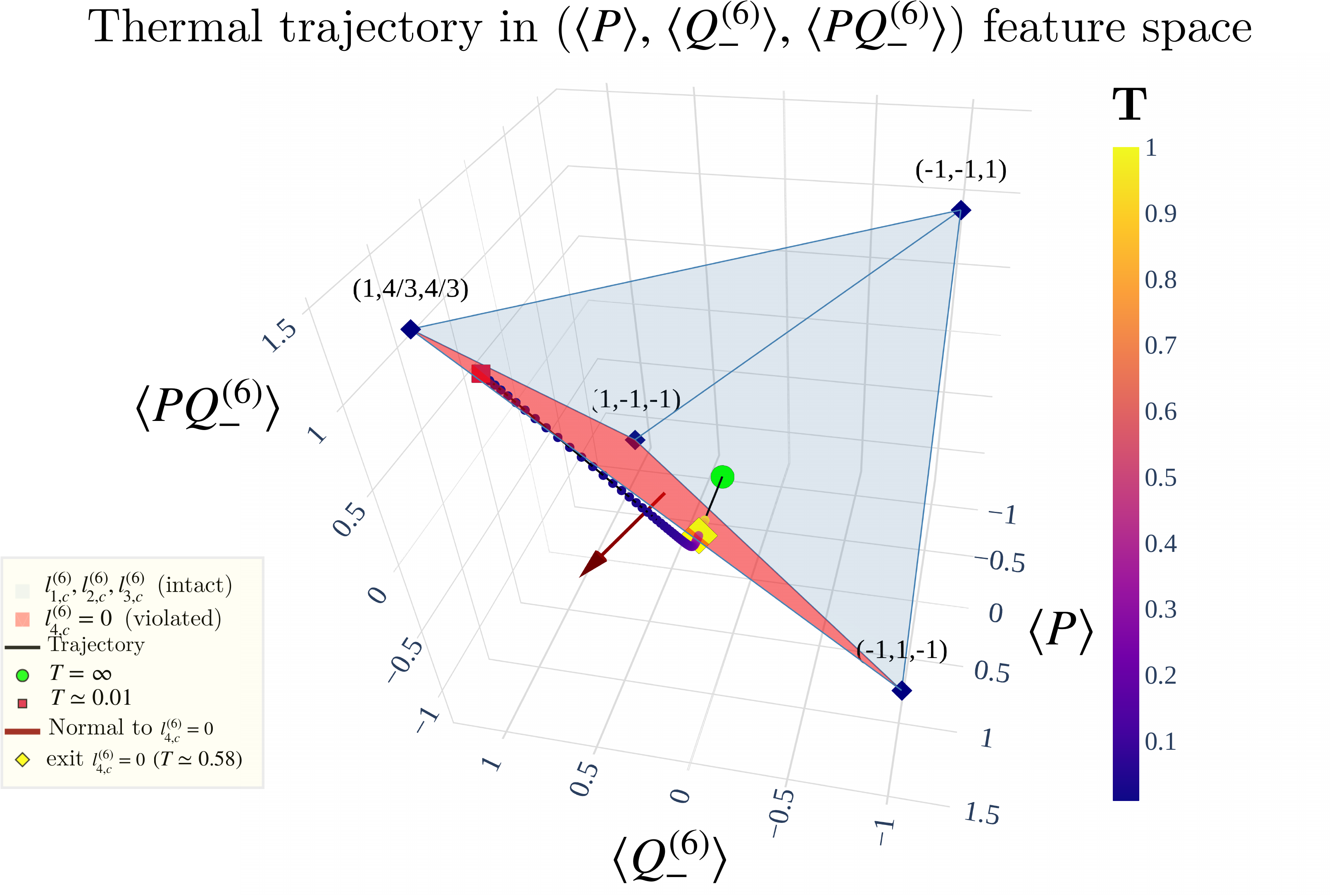}
    }
    \hfill
    \subfloat[Facet $l_{+}^{(6)}$: $P=+1$ sector
    \label{fig:6site-trajectory.facet81}]{
        \includegraphics[width=0.48\textwidth]{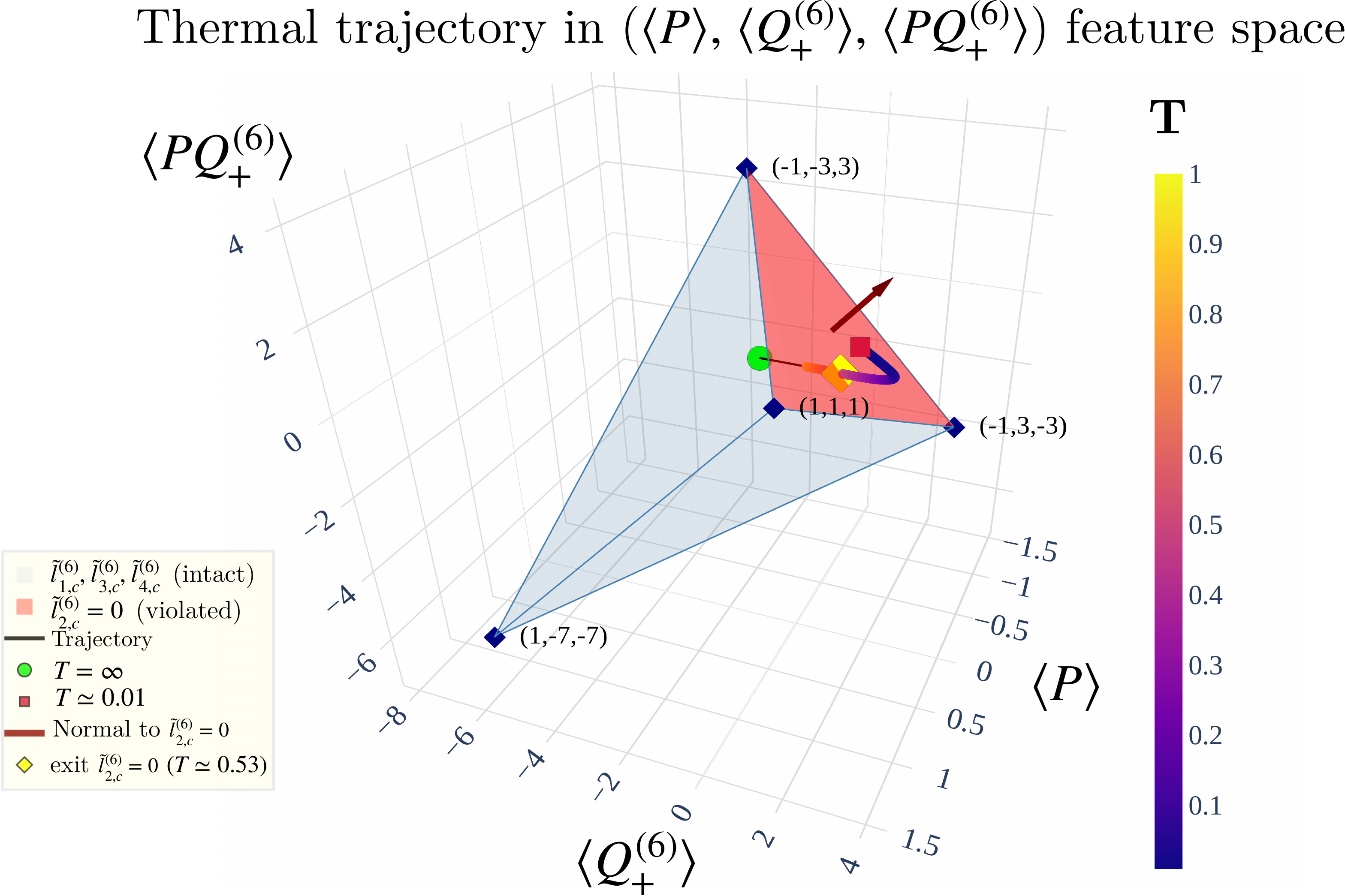}
    }
    \quad
     \subfloat[Temperature dependence of the facet functions \label{fig:facet_plus_minus.2Dtrajectory}]{
        \includegraphics[width=0.38\textwidth]{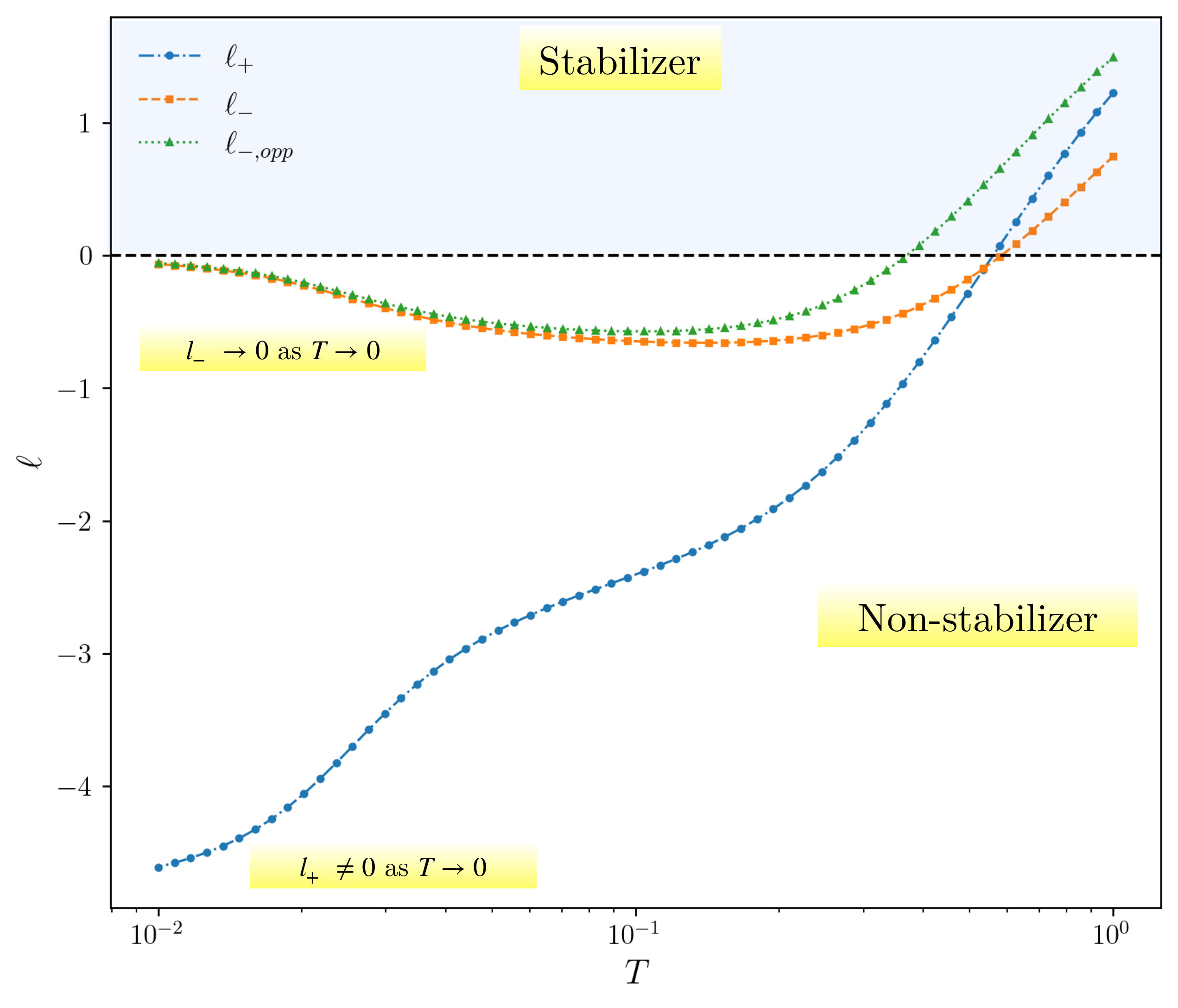}
    }
    \qquad
         \subfloat[Thermal trajectory in the $(l_{-}^{(6)},l_{+}^{(6)})$ plane \label{fig:facetplus_minus.1Dtrajectory}]{
        \includegraphics[width=0.38\textwidth]{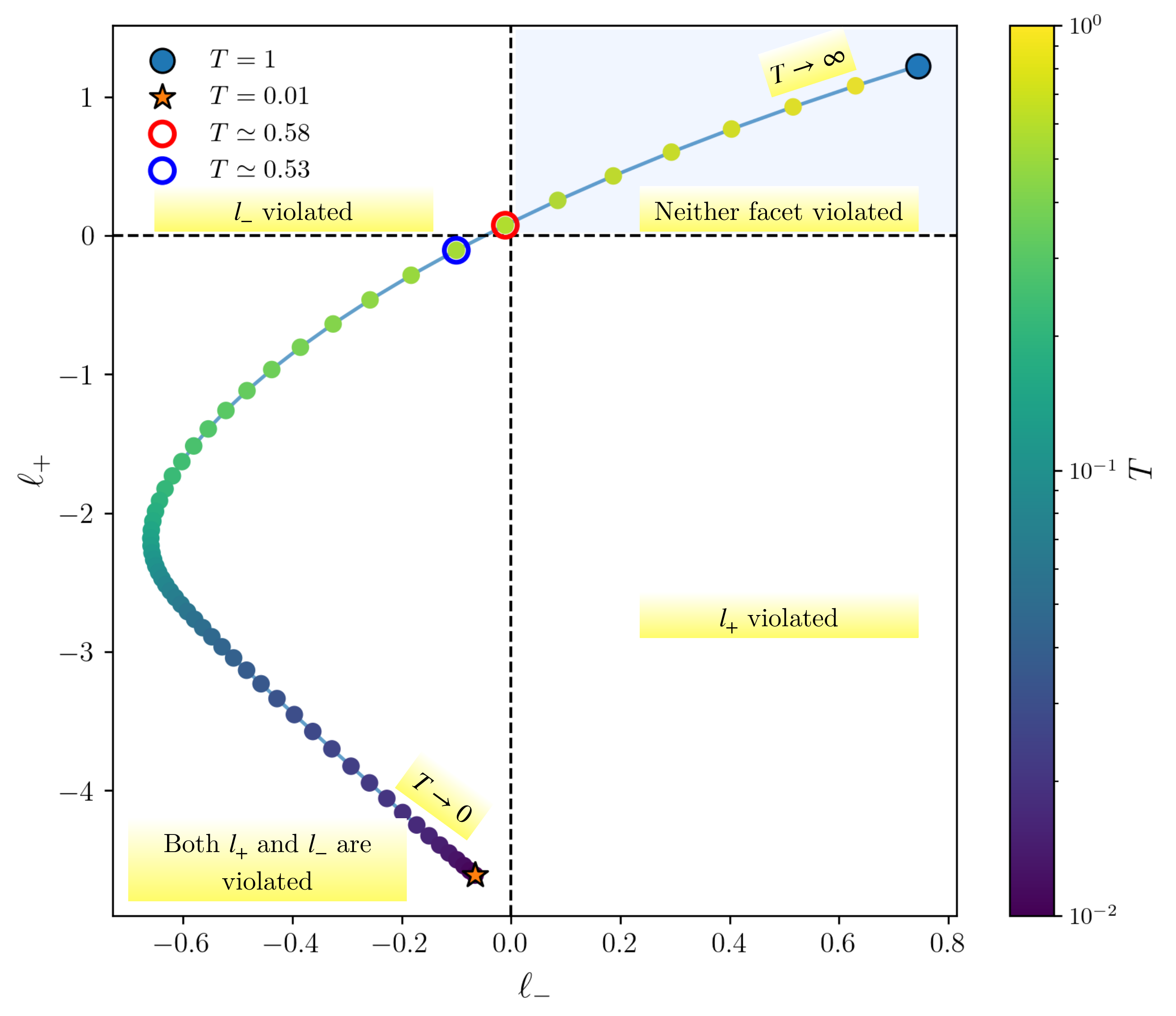}
    }
    \caption{
    Sector-resolved thermal evolution of the six-site marginal.
    (a) Thermal trajectory of $\rho^{(6)}(T)$ in the compressed feature space
    $(\langle P\rangle,\langle Q_{-}^{(6)}\rangle,\langle P Q_{-}^{(6)}\rangle)$ associated with facet $l_{-}^{(6)}$. The highlighted red face corresponds to the $P=-1$-resolved facet, and the yellow diamond marks the point at $T\simeq0.58$ where the trajectory reaches this boundary and subsequently exits the projected stabilizer polytope.
    (b) The same thermal trajectory in the compressed feature space
    $(\langle P\rangle,\langle  Q_{+}^{(6)}\rangle,\langle P Q_{+}^{(6)}\rangle)$ associated with facet $l_{+}^{(6)}$. Here the red face resolves correlations in the complementary $P=+1$ sector, with the corresponding crossing occurring at $T\simeq0.56$.
    (c) Temperature dependence of the two sector-resolved facet functions. The onset facet $l_-^{(6)}$ is violated first, followed by $l_+^{(6)}$ and $l_{-,\mathrm{opp}}^{(6)}$. The violation of $P = -1$ facets weakens toward lower temperature, while the $P = +1$ facet remains violated.
    (d) The same thermal evolution represented directly in the plane of the two facet functions, with the color indicating temperature. The marked points indicate the two facet crossings at $T\simeq0.58$ and $T\simeq0.56$, while the different regions distinguish where neither, one, or both facets are violated. Together, the panels show the evolution from an initial $P=-1$-resolved violation to a low-temperature regime dominated by correlations resolved within the $P=+1$ sector.
    }\label{fig:6site-trajectories}
\end{figure*}
We now turn to the the six-site hexagonal plaquette, $\rho^{(6)}$. This is of particular interest, since local multipartite entanglement in the KHM has been shown to be supported on closed cycles, both in the ground-state \cite{Lyu2026} and at finite-temperature \cite{Sabharwal2025characterizing}. It is therefore natural to ask whether the magic contained in the same local geometry exhibits a structured organization. We find that the six-site magic witness displays a richer version of the structure observed for the three and four-site marginals.
The witness, as before, can be organized into sector-resolved and dressed combinations of composite operators. However, the organizing structure is now provided by the local bond algebra \cite{CobaneraOrtizNussinov2011, NussinovOrtiz2009} and the symmetries of the hexagon.\\

\begin{figure}[t!]
    \centering
    \includegraphics[width=1.\linewidth]{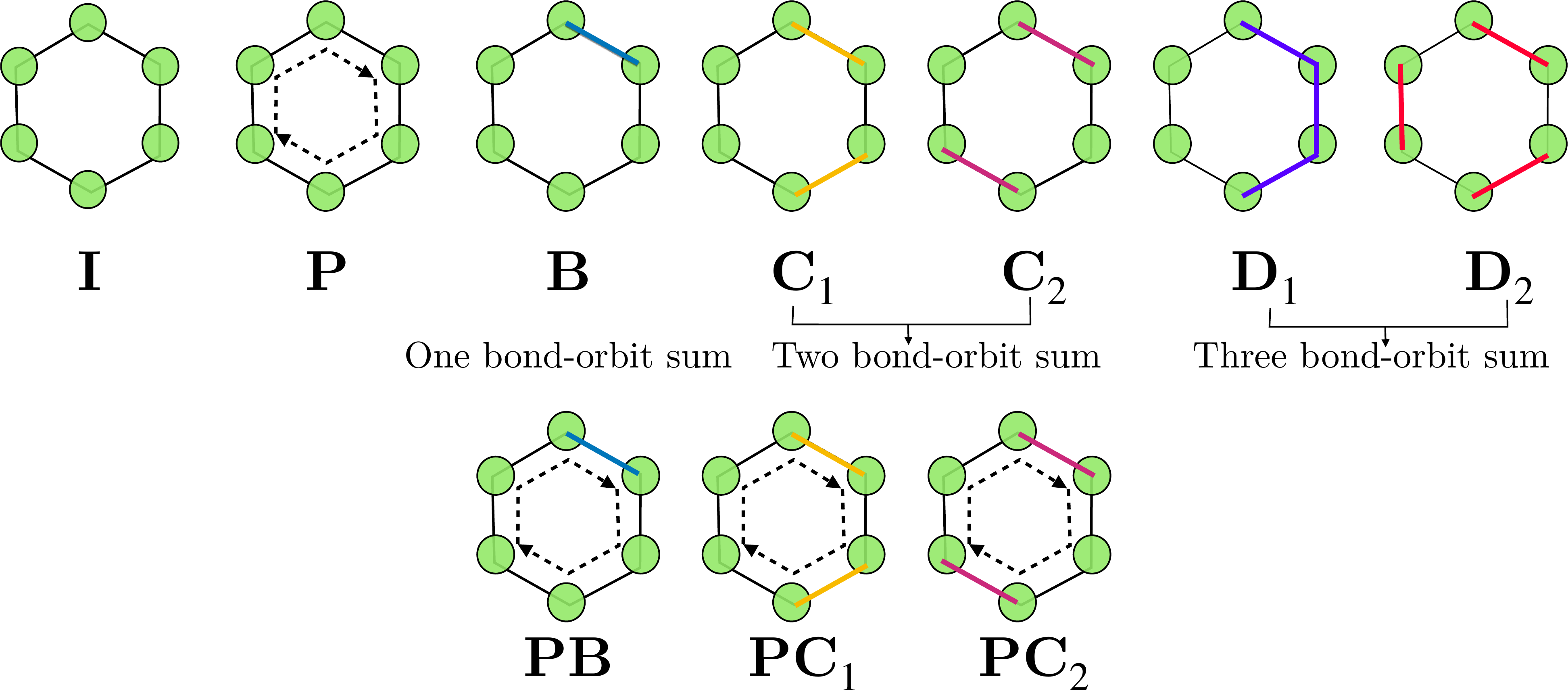}
    \caption{Schematic representation of the operators spanning the six-site marginal. Colored bonds indicate representative terms contributing to $B, C_1, C_2, D_1, D_2,$ while the plaquette operator $P$ is represented by the closed cycle around the hexagon. The $P$-dressed operators $PB, PC_1, PC_2$ are shown by combining this plaquette cycle with the corresponding representative bond pattern.}
    \label{fig:6site.operator.space.basis}
\end{figure}

To see this, let us introduce the six bond operators contained within the hexagonal marginal,
\begin{align}
    K_{1}&=ZZIIII,\quad K_2 = IYYIII ,\quad K_3 = IIXXII \nonumber\\
    K_{4}&= IIIZZI, \quad K_{5} = IIIIYY, \quad K_6 = XIIIIX
\end{align}
The bond operators obey the local commutation relations
\begin{align}
    K_iK_j = (-1)^{\eta_{ij}}K_jK_i
\end{align}
where
\begin{equation}
    \eta_{ij}=
    \begin{cases}
        1, & j \equiv i\pm1 \pmod{6},\\
        0, & \text{otherwise}.
    \end{cases}
\end{equation}
Thus, two neighboring bond operators anticommute, whereas non-neighboring bonds commute. For example,
\begin{equation}
    \{K_1, K_2 \} = 0,\qquad [K_1, K_3] = 0
\end{equation}
The product of these gives the plaquette operator
\begin{equation}
    P = K_1K_2K_3K_4K_5K_6
\end{equation}
which commutes with the bond operators
\begin{equation}\label{eq:P.commute.K}
[P,K_i]=0,
\qquad i=1,\ldots,6\;.
\end{equation}
Motivated by this link-generated algebra, we organize the Pauli terms of the witness into symmetry-adapted combinations of products of the $K_i$. The relations among the terms are made explicit by the transformation $g$, which combines a cyclic rotation of one-site around the hexagon with the spin relabeling
\begin{equation}
    X \longmapsto Z,\quad Z \longmapsto Y,\quad Y \longmapsto X\;.
\end{equation}
Thus, under this transformation, the bond operators are cyclically permuted
\begin{equation}
    K_1 \overset{g}{\longmapsto} K_2 \overset{g}{\longmapsto} K_3 \overset{g}{\longmapsto} K_4 \overset{g}{\longmapsto} K_5 \overset{g}{\longmapsto} K_6 \overset{g}{\longmapsto} K_1
\end{equation}
while the plaquette operator remains invariant
\begin{equation}
    g(P) = P\;.
\end{equation}
Next, we define the orbit of an operator $A$ under $g$
\begin{equation}
    \mathcal{O}_{g}(A) = \{A, g(A), g^2(A),\ldots \}\;,
\end{equation}
where the sequence terminates when the original operator is recovered. We then associate with each orbit its symmetry-adapted sum,
\begin{equation}
    \mathcal{S}_{g}(A) = \sum_{A' \in \mathcal{O}_{g}(A)}A'\;.
\end{equation}
The composite operators entering the six-site witness are obtained by applying this construction to products of one, two, and three link operators, Fig.~\ref{fig:6site.operator.space.basis}. Starting from a single bond, repeated application of $g$ generates all six bonds around the hexagon, and their orbit sum defines
\begin{subequations}
\begin{align}\label{eq:bond.operator}
    B &= \mathcal{S}_{g}(K_1) \\
      &= K_1 + K_2 + K_3 + K_4 + K_5 + K_6   
\end{align}
\end{subequations}
Applying the same construction to the product of two bonds separated by one edge, $K_1K_3$, gives
\begin{subequations}
\begin{align}\label{eq:C1.operator}
    C_1 &= \mathcal{S}_{g}(K_1K_3) \\ 
        &= K_1K_3 + K_2K_4 + K_3K_5 +K_4K_6 + K_5K_1 + K_6 K_2\;,
\end{align}
\end{subequations}
and, pair of opposite bond, $K_1K_4$, gives,
\begin{subequations}
\begin{align}\label{eq:C2.operator}
C_2 &= \mathcal{S}_{g}(K_1K_4) \\
&= K_1K_4 + K_2K_5 + K_3K_6\;.
\end{align}
\end{subequations}
Here we have used the fact that $g$ preserves operator products,
\begin{equation}
g(K_iK_j)=g(K_i)g(K_j),
\end{equation}
so that, for example,
\begin{equation}
g(K_1K_3)=K_2K_4.
\end{equation}
In addition to the one and two-bond orbit sums, there are two inequivalent configurations involving three bonds. The first consists of three consecutive bonds
\begin{subequations}
\begin{align}\label{eq:D1.operator}
    D_1 &= \mathcal{S}_{g}(-K_1K_2K_3)\\
        &=-(K_1K_2K_3 +K_2K_3K_4 +K_3K_4K_5 \nonumber\\
        &\quad\;+K_4K_5K_6+K_5K_6K_1+K_6K_1K_2).
\end{align}
\end{subequations}
The other configuration consists of three alternating bonds, with the orbit sum being
\begin{subequations}
\begin{align}\label{eq:D2.operator}
    D_2 &= \mathcal{S}_{g}(K_1K_3K_5) \\
        &= K_1K_3K_5 + K_2K_4K_6\;.
\end{align}
\end{subequations}
Since the plaquette operator $P$ commutes with every $K_i$ Eq.~\ref{eq:P.commute.K}, it also commutes with their products and with the corresponding orbit sums. This allows us to introduce the plaquette-dressed operators
\begin{equation}
PB,\qquad PC_1,\qquad PC_2.
\end{equation}
Because $P$ is invariant under $g$, these are themselves symmetry-adapted orbit sums. For example,
\begin{equation}
PB = P\mathcal{S}_g(K_1) = \mathcal{S}_g(PK_1),
\end{equation}
and similarly,
\begin{equation}
PC_1 = \mathcal{S}_g(PK_1K_3), \qquad PC_2 = \mathcal{S}_g(PK_1K_4).
\end{equation}
The undressed and dressed operators parallels the structure found for the four-site marginals,
\begin{equation}
(S^{(n)}, \Gamma_a^{(n)}, S^{(n)}\Gamma_a^{(n)}) \quad\longleftrightarrow\quad (P, O, PO)\;.
\end{equation}
with  $O\in{B,C_1,C_2, D_1, D_2}$. Here, the plaquette operator $P$ plays the role of the sector operator $S^{(n)}$, while $O$ and $PO$ provide the corresponding undressed and dressed correlations. No additional dressed partners are required for $D_1$ and $D_2$, since multiplication by $P$ only permutes the terms within their respective orbits:
\begin{equation}\label{eq:PD_i=D_i}
PD_1=D_1, \qquad PD_2=D_2.
\end{equation}
With the composite operators in place, we define the ten-dimensional real operator space
\begin{equation}\label{eq:six.Operator.space}
\mathcal{V}^{(6)}=\text{span}_{\mathbb{R}}
\{I,P, B, C_1, C_2, D_1, D_2, PB, PC_1, PC_2\}\;,
\end{equation}
 visually illustrated in Fig.~\ref{fig:6site.operator.space.basis}.
The basis operators are mutually orthogonal with respect to the HS inner-product Eq.~\ref{eq:HSproduct}
\begin{subequations}
\begin{align}
\Tr(O_aO_b) &= \Tr(O_a^2)\delta_{ab} = 2^6 N_a\delta_{ab}\\
            &= 2^6 \text{diag}(1,1,6,6,3,6,2,6,6,3),
\end{align}
\end{subequations}
where $N_a$ is the  the number of distinct Pauli strings in $O_a$ and 
\begin{equation}\label{eq:basis.six.site}
    O_a\in \{I,P,B,C_1,C_2,,D_1,D_2,PB,PC_1,PC_2\}\;.
\end{equation} 
In addition to being mutually orthogonal, this operator space, much like in the three and four-site marginals, is closed under the Jordan product Eq.~\ref{eq:JordanProduct}. Together with the HS inner-product Eq.~\ref{eq:HSproduct}, this closure equips $\mathcal{V}^{(6)}$ with the structure of a ten-dimensional EJA. We refer to this structure as the \emph{symmetry-resolved plaquette algebra}. The detailed algebraic structure is discussed in Appendix.~\ref{Append:sec.algebra}. \\

Having identified this compact operator space, we find the optimal witness at the onset temperature takes the form
\begin{align}
    W^{(6)} &= \Pi_{+}^{(P)} + \Pi_{-}^{(P)}\bigg(\frac{1}{2}I+\frac{1}{3}B-\frac{1}{6}C_1-\frac{1}{6}C_2\bigg) \\
    \Pi_{\pm}^{(P)} &= \frac{I \pm P}{2}\;,
\end{align}
where the witness acts as the identity in the $P=+1$ sector, while in the $P=-1$ sector it probes a structured combination of one and two-bond orbit operators. Thus, the witness is not an arbitrary sum of Pauli strings, but is entirely supported on the same symmetry-adapted composite operators introduced above. This motivates constructing the projected stabilizer polytope in the corresponding feature coordinates:
\begin{align}
    \mathbf{x}^{(6)}(\omega) = \bigg( &\langle P \rangle_{\omega}, \langle B\rangle_{\omega},\langle C_1\rangle_{\omega},\langle C_2\rangle_{\omega},\langle D_1\rangle_{\omega},\langle D_2 \rangle_{\omega},\nonumber \\ 
    &\langle PB\rangle_{\omega},\langle PC_1\rangle_{\omega},\langle PC_2 \rangle_{\omega}\bigg)\;.
\end{align}
We next obtain the facet inequalities, satisfied by every stabilizer state, within this feature space. The resulting polytope has $15$ vertices and $18$ facets, whose inequalities are listed in Table~\ref{tab:six.site.facets}. Of these, $7$ are trivial and $11$ are non-trivial. Interestingly, only three of these inequalities are violated along the thermal trajectory,
\begin{align}\label{eq:91.six.site.projector}
l_{-}^{(6)}[\omega]
&=
\left\langle
\Pi_-^{(P)}
\left(
6I-4B+2C_1+2C_2
\right)
\right\rangle_{\omega}
\geq 0,
\end{align}
\begin{align}\label{eq:81.six.site.projector}
l_{+}^{(6)}[\omega]
&=
\left\langle
\Pi_+^{(P)}
\left(
6I-2B-2C_1-2C_2-D_1+3D_2
\right)
\right\rangle_{\omega}
\geq 0\;.
\end{align}
\begin{align}
    l_{-,\mathrm{opp}}^{(6)}[\omega] = \left\langle \Pi_-^{(P)}(6I-2B-2 C_2) \right\rangle_\omega \geq 0\;,
\end{align}
We highlight two facets $l_{-}^{(6)}$  and $l_{+}^{(6)}$that become particularly relevant along the thermal trajectory. The facet, $l_{-}^{(6)}$, which is the first of these two facets crossed by the thermal trajectory, reproduces the symmetry-adapted witness at the onset of magic
\begin{equation}
    \ell[\omega] = \frac{l^{(6)}_{-}[\omega]}{12}\;.
\end{equation}
As in the three and four-site cases, the structure of the onset witness allows us to further identify a witness-adapted three-dimensional projection of the full feature space. Defining
\begin{equation}\label{eq:Q.six.site}
Q_{-}^{(6)} =\frac{1}{2}I +\frac{1}{3}B -\frac{1}{6}C_1 -\frac{1}{6}C_2\;,
\end{equation}
the symmetry-adapted witness can be written compactly as
\begin{equation}\label{eq:Witness.six.site.compact}
W^{(6)} = \Pi_+^{(P)} + \Pi_-^{(P)}Q_{-}^{(6)}.
\end{equation}
Since $P$ commutes with $B$, $C_1$, and $C_2$, it also commutes with $Q^{(6)}$,
\begin{equation}
    [P, Q_{-}^{(6)}] = 0\;.
\end{equation}
Consequently, we define the compressed feature vector
\begin{equation}\label{eq:six.site.compr.feature.vector}
\mathbf{x}_{c}^{(6)}(\omega)=(\langle P\rangle_{\omega},\langle Q_{-}^{(6)}\rangle_{\omega}, \langle PQ_{-}^{(6)}\rangle_{\omega})\;.
\end{equation}
wherein the polytope is again a tetrahedron, specified by the four facet inequalities:
\begin{subequations}
\begin{align}
l_{1,c}^{(6)}[\omega] &= 1 +\langle P\rangle_{\omega} +\langle Q_{-}^{(6)}\rangle_{\omega} +\langle PQ_{-}^{(6)}\rangle_{\omega} \geq 0,\\
l_{2,c}^{(6)}[\omega] &=
4
+4\langle P\rangle_{\omega}
-3\langle Q_{-}^{(6)}\rangle_{\omega}
-3\langle PQ_{-}^{(6)}\rangle_{\omega}
\geq 0,
\\
l_{3,c}^{(6)}[\omega]&=
1
-\langle P\rangle_{\omega}
+\langle Q_{-}^{(6)}\rangle_{\omega}
-\langle PQ_{-}^{(6)}\rangle_{\omega}
\geq 0,
\\
l_{4,c}^{(6)}[\omega]&=
1
-\langle P\rangle_{\omega}
-\langle Q_{-}^{(6)}\rangle_{\omega}
+\langle PQ_{-}^{(6)}\rangle_{\omega}
\geq 0.
\end{align}
\end{subequations}
The final inequality is the compressed representation of the onset facet $l_{-}^{(6)}$
\begin{equation}
\ell[\omega]
= \frac{1}{2}l_{4,c}^{(6)}[\omega]\;.
\end{equation}
A similar compression can be constructed for the facet $l_{+}^{(6)}$. In this case, we define
\begin{equation}
    Q_{+}^{(6)} = -I + \frac{2(B + C_1 + C_2)}{3} + \frac{D_1}{3} - D_2 \;,
\end{equation}
in which case, the witness can be compactly written as
\begin{equation}\label{eq:}
    \tilde{W}^{(6)} = \Pi_-^{(P)} + \Pi_+^{(P)}Q_{+}^{(6)}.
\end{equation}
In contrast to the onset witness, the sector structure is reversed, with $Q_{+}^{(6)}$ acting in the $P=+1$ sector rather than the $P=-1$ sector. Moreover, we can similarly obtain a compressed feature space by choosing the following
\begin{equation}
\tilde{\mathbf{x}}_{c}^{(6)}(\omega)=(\langle P\rangle_{\omega},\langle Q_{+}^{(6)}\rangle_{\omega}, \langle PQ_{+}^{(6)}\rangle_{\omega})\;.
\end{equation}
In this feature space, the projected polytope is a tetrahedron, specified according to the following inequalities:
\begin{subequations}
\begin{align}
\tilde{l}_{1,c}^{(6)}[\omega] &= 7 + 7\langle P\rangle_{\omega} +\langle Q_{+}^{(6)}\rangle_{\omega} +\langle PQ_{+}^{(6)}\rangle_{\omega} \geq 0,\\
\tilde{l}_{2,c}^{(6)}[\omega] &=
1
+\langle P\rangle_{\omega}
-\langle Q_{+}^{(6)}\rangle_{\omega}
-\langle PQ_{+}^{(6)}\rangle_{\omega}
\geq 0,
\\
\tilde{l}_{3,c}^{(6)}[\omega]&=
3
-3\langle P\rangle_{\omega}
+\langle Q_{+}^{(6)}\rangle_{\omega}
-\langle PQ_{+}^{(6)}\rangle_{\omega}
\geq 0,
\\
\tilde{l}_{4,c}^{(6)}[\omega]&=
3
-3\langle P\rangle_{\omega}
-\langle Q_{+}^{(6)}\rangle_{\omega}
+\langle PQ_{+}^{(6)}\rangle_{\omega}
\geq 0\;,
\end{align}
\end{subequations}
where the second inequality is the compressed representation of the facet $l^{(6)}_{+}$
\begin{equation}
    \tilde{\ell}[\omega] = \frac{1}{2}\tilde{l}_{2,c}^{(6)}\;.
\end{equation}
Now, much like in the previous two cases, the onset of magic corresponds to the thermal trajectory reaching the boundary facet contained in the supporting hyperplane
\begin{equation}\label{eq:91.six.site.hyperplane}
    l_{4,c}^{(6)} = 0\;,
\end{equation}
and subsequently, leaving the polytope at $T \simeq 0.58$, as shown in Fig.~\ref{fig:6site-trajectory.facet91}. Although this facet accurately captures the first emergence of magic, its violation weakens again upon further cooling, with $l^{(6)}_{4,c}$ approaching zero in the low-temperature limit, Fig.~\ref{fig:facet_plus_minus.2Dtrajectory}. In contrast, the thermal trajectory in the $\big(\langle P\rangle,\langle Q_{+}^{(6)}\rangle,\langle PQ_{+}^{(6)}\rangle\big)$ feature space, Fig.~\ref{fig:6site-trajectory.facet81}, after reaching the boundary facet contained in the hyperplane
\begin{equation}\label{eq:81.six.site.hyperplane}
    \tilde{l}_{2,c}^{(6)} = 0    
\end{equation}
 leaves the tetrahedron polytope at a slightly lower temperature $T \simeq 0.56$, and yet remains relevant even in the low-temperature regime, Fig.~\ref{fig:facet_plus_minus.2Dtrajectory}. The two facets therefore probe different parts of the thermal evolution. While $l_{-}^{(6)}$ provides the earliest signature of local magic, $l_{+}^{(6)}$ more faithfully captures its persistence toward low temperature. 
 
 Unlike the onset facet $l_-^{(6)}$, the facet $l_{-,\mathrm{opp}}^{(6)}$ is first violated at a lower temperature, $T\simeq0.37$. Thus, although three distinct facets are crossed along the thermal trajectory, the two previously highlighted facets provide the primary characterization of the onset and subsequent evolution of local magic.

The earlier violation of $l_{-}^{(6)}$ can be understood from its sector-resolved form. Since, $\langle P\rangle\simeq 0$ at the onset, the two plaquette sectors carry approximately equal weight. The crossing of $l_{-}^{(6)}$ therefore does not arise from an enhanced occupation of the $P=-1$ sector, but from the correlations resolved within it. In particular, the nearest-neighbor bond correlations contained in $B$, 
develop more rapidly upon cooling than the higher-order correlations $C_1$ and $C_2$ \footnote{As shown below, the remaining composite operators spanning the operator space arise naturally from the operator hierarchy generated by powers of $B$.}. Consequently, the $-4B$ contribution in Eq.~\ref{eq:91.six.site.projector} drives the facet toward violation before the compensating contributions from $C_1$ and $C_2$ become sufficiently large. By contrast, the facet $l_{+}^{(6)}$ in Eq.~\ref{eq:81.six.site.projector} probes the complementary $P=+1$ sector and contains a broader set of higher-order correlations, including $C_1$, $C_2$, $D_1$, and $D_2$. It therefore becomes violated at a slightly lower temperature, once these multi-bond correlations have developed more strongly. The two crossings therefore distinguish two stages in the development of local magic: an initial violation controlled by the $P=-1$-resolved bond sector, where the rapid growth of the nearest-neighbor correlations in $B$ dominates over $C_1$ and $C_2$, followed by a lower-temperature regime in which the complementary $P=+1$ sector and the higher-order correlations $C_1,C_2,D_1$, and $D_2$ become increasingly important and saturates magic in
the ground state.

\subsection{Symmetry origin of the operator space and exact reduced robustness of magic}\label{subsec:symmetry.exactness}
The operator space introduced above can also be understood independently of the structure of the optimal witness. For the isotropic Kitaev model, the operators that can acquire a non-zero expectation value on the hexagonal marginal are constrained by flux conservation~\cite{Kitaev2006}, the combined sixfold lattice--spin symmetry $\mathcal{C}_6$, and time-reversal-even symmetry $\mathcal T$.\\

Starting from the $4^6=4096$ Pauli operators on six spins, flux conservation \cite{Kitaev2006} restricts the allowed operators to those which commute with the conserved plaquette fluxes. On the hexagon, these form a $64$-dimensional algebra generated by the six bond operators $K_i$, closely related to the bond-algebraic formulation developed in \cite{NussinovOrtiz2009, CobaneraOrtizNussinov2011}. The combined $\mathcal{C}_6$ symmetry further groups these operators to $14$ independent Hermitian operators, and time-reversal-even symmetry removes four additional operators. We therefore obtain
\begin{equation}
    4096 \xrightarrow{\mathrm{flux}} 64 \xrightarrow{\mathcal{C}_6} 14 \xrightarrow{\mathcal T} 10.
\end{equation}
The resulting $10$ dimensional space is precisely our operator space Eq.~\ref{eq:six.Operator.space}. Thus, for a state that exactly respects these symmetries, the restriction to $\mathcal V^{(6)}$ is not an approximation. The state itself lies in this operator space.\\

To see the consequence of this for magic, let $\mathcal{E}_{\mathcal{V}^{(6)}}$ or  simply $\mathcal{E}_{\mathcal{V}}$,  denote a symmetry projection. For a state invariant under the symmetries, we have
\begin{equation}
    \mathcal{E}_{\mathcal{V}}(\rho^{(6)}) = \rho^{(6)} \;.
\end{equation}
Importantly, $\mathcal{E}_{\mathcal{V}}$ also maps the stabilizer polytope into itself. The flux projection is an average over Pauli operations, while the $\mathcal{C}_6$ projection is an average over Clifford operations. In either case, a pure stabilizer state is mapped to another pure stabilizer state. Time reversal similarly maps stabilizer states to stabilizer states. Hence, 
\begin{equation}
    \mathcal{E}_{\mathcal V}(\mathrm{STAB}_6)\subseteq\mathrm{STAB}_6.
\end{equation}
In this sense, $\mathcal{E}_{\mathcal{V}}$ is stabilizer preserving. It cannot generate magic from a stabilizer state. This observation leads us to a more general statement.\\

Let $\mathcal{F}$ denote a convex set of free states and let $\mathcal{E}$ be the projection satisfying 
\begin{equation}\label{eq:free.state.preserve}
  \mathcal{F}_{\mathcal{E}} \equiv  \mathcal{E}(\mathcal{F}) \subseteq \mathcal{F}\;.
\end{equation}
where $\mathcal{F}_{\mathcal{E}}$ is our reduced free set. 
Then for a state fixed by the projection
\begin{equation}\label{eq:Projection}
    \mathcal{E}(\rho) = \rho\;,
\end{equation}
the robustness computed with respect to $\mathcal{F}_{\mathcal{E}}$ is equal to the robustness in the full free set $\mathcal{F}$
\begin{equation}\label{eq:redRom.equal.RoM}
   \mathcal{R}_{\mathcal{E}}[\rho]= \mathcal{R}[\rho]
\end{equation}
This result follows from the two directions of the optimization. First, we consider an optimal decomposition in $\mathcal{F}$,
\begin{equation}
    \rho = (1 + r^*) \sigma_{+} - r^*\sigma_{-},\quad \sigma_{\pm} \in \mathcal{F}\;.
\end{equation}
Applying, $\mathcal{E}$ and using Eq.~\ref{eq:Projection} gives
\begin{equation}
\rho=(1+r^{*})\mathcal{E}(\sigma_+)-r^{*}\mathcal{E}(\sigma_-)\;.
\end{equation}
Since $\mathcal{E}(\sigma_\pm)\in\mathcal F_{\mathcal{E}}$, this gives us an allowed reduced decomposition with exactly the same value of $r^*$. The reduced optimization can therefore only return the same or a smaller minimum,
\begin{equation}
\mathcal R_\mathcal{E}[\rho]\leq \mathcal R[\rho]\;.
\end{equation}
This is the usual lower-bound direction of a reduced RoM construction \cite{Varela2026}. The reverse direction follows from the free-state preserving property of the projection, Eq.~\ref{eq:free.state.preserve}. Let us now consider instead an optimal reduced decomposition,
\begin{equation}
\rho=(1+r_{\mathcal{E}}^*)\tau_+-r_{\mathcal{E}}^*\tau_-,\qquad \tau_\pm\in\mathcal F_{\mathcal{E}}\;.
\end{equation}
Owing to Eq.~\ref{eq:free.state.preserve}, the same $\tau_\pm$ are also allowed free states in the full optimization.
Thus, any decomposition available to the reduced problem is therefore also available to the full problem. Since the full problem minimizes over a larger set of possible decompositions, its minimum cannot be larger,
\begin{equation}
\mathcal R[\rho]\leq\mathcal R_{\mathcal{E}}[\rho]\;.
\end{equation}
Together,
\begin{equation}
\mathcal R_{\mathcal{E}}[\rho]\leq\mathcal R[\rho]\leq\mathcal R_{\mathcal{E}}[\rho]\;,
\end{equation}
thus, leads to Eq.~\ref{eq:redRom.equal.RoM}. 
The physical content of this result is simple. Although the full optimization has access to more free-state decompositions, this additional freedom does not provide an advantage for a state fixed by a free-state preserving projection. Any full-space decomposition can be projected into the reduced space without changing its cost, while every reduced decomposition remains valid in the full problem.\\

For the six-site Kitaev marginal, we take
\begin{equation}
\mathcal F=\mathrm{STAB}_6,\qquad \mathcal{E}={\mathcal{E}}_{\mathcal V}\;.
\end{equation}
Consequently, whenever the marginal respects the flux, $\mathcal{C}_6$, and time-reversal symmetries $\mathcal{T}$ exactly,
\begin{equation}
 \mathcal{R}_{\mathcal V}[\rho^{(6)}] \equiv  \mathcal{R}_{\mathcal E_{\mathcal{V}}}[\rho^{(6)}]=\mathcal{R}[\rho^{(6)}]\;.
\end{equation}

The reduction therefore preserves more than the onset of magic. It preserves the complete value of the robustness, with faithfulness following immediately,
\begin{equation}
\mathcal R_{\mathcal{V}}[\rho^{(6)}]>1\quad\Longleftrightarrow\quad\mathcal R[\rho^{(6)}]>1\;,
\end{equation}
with both quantities taking the same value throughout the symmetry-fixed family. This is especially useful here because normalization fixes the identity component, leaving only the nine non-trivial expectation values
\begin{equation}
\langle P\rangle,\ \langle B\rangle,\ \langle C_1\rangle,\ \langle C_2\rangle,\ \langle D_1\rangle,\ \langle D_2\rangle,\ \langle PB\rangle,\ \langle PC_1\rangle,\ \langle PC_2\rangle
\end{equation}
to specify the state within this sector. The symmetry reduction can therefore strongly reduce both the number of quantities that need to be determined and the complexity of the stabilizer optimization, without sacrificing the value of the RoM.\\

For the rdms numerically considered in the following, the exact statement applies, provided that these symmetries are realized exactly. Any small component outside $\mathcal V^{(6)}$, for example due to finite-size or TPQ sampling effects, instead leads to
\begin{equation}
\mathcal{E}_{\mathcal V}(\rho)\neq\rho\;,
\end{equation}
in which case the reduced RoM retains its lower-bound character. We return to this point below by quantifying how accurately the rdms, obtained using the TPQ method, are captured by the symmetry-adapted operator space. 
\subsection{How much information is captured by the operator space?}
\label{subsec:thermal.state}

The preceding discussion shows that the ten-dimensional space $\mathcal V^{(6)}$ is fixed by the symmetries of the isotropic KHM and provides an exact description whenever these symmetries are realized exactly. We now ask how this structure appears in the finite-temperature rdms obtained numerically, and how much of the state
can already be captured by an even simpler description involving only the dominant bond correlations.\\

We begin with the modular Hamiltonian
\begin{equation}
H_{\text{mod}}^{(6)}= -\log\rho^{(6)}\;.
\end{equation}
Since only the traceless part of an operator affects its nontrivial structure, we remove its trace component and define
\begin{equation}
\widetilde H_{\text{mod}}^{(6)}
=
H_{\text{mod}}^{(6)}
-
\frac{\Tr H_{\text{mod}}^{(6)}}{2^6}I\;.
\end{equation}
We find that at the onset temperature, the decomposition of $\widetilde H_{\text{mod}}^{(6)}$ is strongly dominated by the bond operator,
\begin{equation}
\widetilde H_{\text{mod}}^{(6)} = \lambda B + \delta \widetilde H_{\text{mod}}^{(6)},
\end{equation}
where the coefficients contained in $\delta \widetilde{H}_{\text{mod}}^{(6)}$ are several orders of magnitude smaller than $\lambda$. Here, $\lambda$ is the effective dimensionless bond coupling obtained from the HS projection of the traceless modular Hamiltonian, $\tilde{H}_{\text{mod}}^{(6)}$ onto $B$,
\begin{equation}
\lambda=\frac{\Tr\left[B\widetilde H_{\text{mod}}^{(6)}\right]}{\Tr\left[B^2\right]}\simeq -0.38\;,
\end{equation}
at the onset temperature, $T \simeq 0.58$. This suggests the effective state
\begin{equation}
\rho_B^{(6)} =
\frac{
e^{-\lambda B}
}{
\Tr[e^{-\lambda B}]
}\;.
\end{equation}
Next we expand the exponential term in powers of $\lambda$,
\begin{equation}
e^{-\lambda B} = \sum_{n=0}^{\infty} \frac{(-1)^n\lambda^n}{n!}B^n\;
\end{equation}
which, owing to the algebra of the bond operators, results in
\begin{subequations}
\begin{align}
B^2 &= 6I+2C_1+2C_2,\\
B^3 &= 12B+2D_1+6D_2,\\
B^4 &= 72I+32C_1 + 24C_2+8PC_1,\\
B^5 &= 160B + 40D_1 + 120D_2 + 16PB,\\
B^6 &= 960 I + 96 P + 480 C_1 + 320 C_2 + 192PC_1 + 32PC_2\;.
\end{align}
\end{subequations}
The higher powers of $B$ can themselves be decomposed in the operator basis introduced above. However, these do not generate arbitrary new operator directions. Instead, they reduce to linear combinations of the composite operators contained in the operator space. By organizing these terms according to the lowest power of $\lambda$ at which they appear in the expansion of $e^{-\lambda B}$ makes this structure transparent, as summarized in Table~\ref{tab:bond-exponential-hierarchy}.
\begin{table}[t]
    \centering
    \begin{tabular}{c|c}
        \hline
        Order in $\lambda$ &Operators generated
        \\
        \hline
        $\lambda^{0}$ & $I$ \\
        $\lambda^{1}$ & $B$ \\
        $\lambda^{2}$ & $C_{1},\,C_{2}$ \\
        $\lambda^{3}$ & $D_{1},\,D_{2}$ \\
        $\lambda^{4}$ & $PC_{1}$ \\
        $\lambda^{5}$ & $PB$ \\
        $\lambda^{6}$ & $P,\,PC_{2}$ \\
        \hline
    \end{tabular}
    \caption{Hierarchy of operators generated by the expansion of
    $e^{\lambda B}$. The entries indicate the lowest order in $\lambda$
    at which each operator first appears. Note that the lowest order in $\lambda$ at which each operator first appears coincides with the number of Majorana gauge variables, $u_{ij}^{\alpha}$ entering that operator; see Appendix~\ref{Append:Majorana.operator.space}.}
    \label{tab:bond-exponential-hierarchy}
\end{table}
The composite operators $C_{1},C_{2},D_{1},D_{2},\ldots$ therefore arise naturally from the polynomial expansion of the dominant bond contribution to the modular Hamiltonian.\\

To quantify the accuracy of this effective description, we use the fidelity
\begin{equation}
{F}(\rho,\sigma)
=
\left(
\Tr\sqrt{\sqrt{\rho}\,\sigma\,\sqrt{\rho}}
\right)^2 \; , 
\label{eq:fidelity}
\end{equation}
as plotted in Fig.~\ref{fig:fidelity}.   
At the onset temperature, $T \simeq 0.58$, we find
\begin{equation}
{F}\left(\rho^{(6)},\rho_{B}^{(6)}\right)
\simeq 0.9999.
\end{equation}
Thus, the dominant bond contribution to the modular Hamiltonian provides an exceptionally accurate effective description of the six-site rdm, with a fidelity of approximately $99.99\%$.\\

We next examine the full symmetry-adapted operator space identified above. For the numerical rdm, we construct its projection onto $\mathcal V^{(6)}$,
\begin{equation}
\rho_{\mathcal V}^{(6)} =\sum_a \frac{\Tr\left[O_a\rho^{(6)}\right]}{\Tr\left[O_a^2\right]}O_a,
\end{equation}
where $O_a$ is specified in Eq.~\ref{eq:basis.six.site}. Across the full temperature range, the fidelity of this reconstruction is bounded from below by
\begin{equation}
\min_T
{F}\left(
\rho^{(6)}(T),
\rho_{\mathcal V}^{(6)}(T)
\right)
\simeq 0.9994.
\end{equation}
Thus, even where the approximation performs least accurately, the ten-dimensional operator space reproduces the six-site rdm with a fidelity of approximately $99.94\%$, as shown in Fig.~\ref{fig:fidelity}.
This small infidelity is reflection of the finite accuracy of the numerical methods used to solve the model (thermal pure quantum states, as described in Appendix~\ref{Append:Numerics}).
\\

The same agreement is observed for the three and four-site marginals. Projecting onto their respective three and six-dimensional operator spaces gives
\begin{align}
\min_T {F}\left( \rho^{(3)}(T),
\rho_{\mathcal V}^{(3)}(T)
\right) &\simeq 0.9999,\\
\min_T {F}\left(\rho^{(4)}(T), \rho_{\mathcal V}^{(4)}(T) \right) &\simeq 0.9998.
\end{align}
Thus, the operator spaces identified from the magic witnesses are not restricted to the vicinity of the magic onset, but provide accurate representations of the corresponding local marginals throughout the entire temperature range considered, with fidelities remaining above $99.94\%$ in all three cases.\\

The near-unit fidelities obtained here are therefore not merely an indication that $\mathcal V^{(6)}$ provides a good approximation. For an exactly symmetry-invariant thermal state, the restriction to $\mathcal V^{(6)}$ is exact. The small deviations from unit fidelity observed in the numerical rdms should consequently be attributed to the finite numerical realization of these symmetries, including TPQ
sampling and finite-size effects.

\subsection{Application to genuine multipartite entanglement}
\label{subsec:GME}

The symmetry reduction derived above is not specific to magic.
Since the six-site marginal lies in $\mathcal V^{(6)}$ whenever the relevant symmetries are realized exactly, only the component of an observable within this space can contribute to its expectation
value. This naturally implies that the same compact operator space should also
contain the information required to probe other local quantum
resources. We illustrate this here for genuine multipartite
entanglement (GME) \cite{Sabharwal2025characterizing}.\\

In previous work, we introduced a semidefinite programming (SDP)-based framework for characterizing the depth and spatial structure of entanglement at finite-temperature quantum spin liquids \cite{Sabharwal2025characterizing}. Applying this approach to the KHM, we found that GME develops on the six-site hexagonal plaquette at low temperature.  Through the dual SDP formulation, this optimization also produces a quantitative GME witness tailored to the thermal rdm, which we denote by $W_{\rm GME}$. We project this witness onto the
same symmetry-adapted operator space,
\begin{equation}
W_{\rm GME}^{\mathcal V}
=
\sum_a
\frac{\Tr[O_aW_{\rm GME}]}{\Tr[O_a^2]}O_a\;,
\end{equation}
where
$O_a$ is as specified in Eq.~\ref{eq:basis.six.site}. \\

For a state lying entirely in the symmetry-fixed space $\mathcal V^{(6)}$, components of an observable orthogonal to $\mathcal V^{(6)}$ cannot contribute to its expectation value. Therefore, replacing the full GME witness by its projection onto $\mathcal V^{(6)}$ leaves the expectation value unchanged,
\begin{equation}
\Tr[\rho^{(6)}W_{\rm GME}]=\Tr[\rho^{(6)}W_{\rm GME}^{\mathcal V}]\;.
\end{equation}

For the rdms obtained using TPQ, the two expectation values agree within the numerical accuracy with which the symmetry-fixed space is realized, as shown in Fig.~\ref{fig:GME}.

\section{Discussion}
\label{sec:discussion}
Our results show that the emergence of local magic in the finite-temperature KHM is governed by a highly
structured and unexpectedly small operator space. Starting from the optimal RoM witnesses of the three, four, and six-site marginals, we identified compact families of operators that capture the
thermal evolution of the corresponding reduced states and determine their crossings of the stabilizer boundary. For the six-site hexagon, this structure can be understood directly from the
symmetries of the model. The resulting space $\mathcal V^{(6)}$ is therefore not merely a
numerically successful truncation. For states satisfying these symmetries exactly, the restriction to $\mathcal V^{(6)}$ is exact, and the corresponding reduced RoM is equal to the full RoM.

\subsection{Sector-resolved magic at finite temperature in the  Kitaev spin liquid}
The interpretation of the two facets, [Eq.~\eqref{eq:91.six.site.projector}] and [Eq.~\eqref{eq:81.six.site.projector}] can also be related to the finite-temperature picture of the KHM originally discussed by Nasu \emph{et al.} \cite{Nasu2015}, and subsequent studies thereafter \cite{Yamaji2016, Yoshitake2016, Yoshitake2017_Majorana,Yoshitake2017_Temp, Hermanns2018, Rousochatzakis2019, Li2020, Feng2020, Gohlke2023}. 
In these works, the higher-temperature crossover is associated with the development of nearest-neighbor spin correlations followed by the emergence of the itinerant Majoranas, while the lower-temperature crossover is associated with the thermal evolution of the $\mathbb{Z}_2$ flux. In this regard, the first violation of $l_{-}^{(6)}$ is consistent with the former picture, Fig.~\ref{fig:facet_plus_minus.2Dtrajectory}. 
Although this facet is resolved in the $P=-1$ sector, $\langle P\rangle$ remains close to 
zero at its onset, Fig.~\ref{fig:plaquette}, such that the two plaquette sectors still carry 
approximately equal weight. Its violation is therefore primarily associated with the development 
of the $P=-1$-resolved bond correlations, with the contribution from $B$ becoming important before the higher-order terms $C_1$ and $C_2$.\\

The Majorana representation of these operators provides a more direct interpretation of this hierarchy, as shown in Appendix \ref{Append:Majorana.operator.space}. The plaquette operator, $P$ corresponds to the local $\mathbb{Z}_2$ flux, while $B$ composed of nearest-neighbor matter-Majorana bilinears. In contrast, $C_1$ and $C_2$ correspond to gauge-dressed four matter-Majorana correlations. The first facet can therefore be viewed as resolving the competition between the rapidly developing bilinear matter-Majorana correlations and the corresponding four matter-Majorana correlations within the $P = -1$ sector. This provides a microscopic connection between the onset of local magic and the itinerant-Majorana regime associated with the higher-temperature crossover.\\

The facet $l_{+}^{(6)}$, on the other hand, resolves the complementary $P=+1$ sector and contains the additional operators $D_1$ and $D_2$. However, their Majorana structure is qualitatively different. Although, $D_1$ is a three-bond composite operator, it reduces to a matter-Majorana bilinear connecting opposite sites of the hexagon and dressed by a three-link gauge string. On the other hand, $D_2$ contains a gauge-dressed six matter-Majorana structure. Moreover, owing to Eq.~\ref{eq:PD_i=D_i}, these structures are supported entirely within the $P = +1$ sector. The onset of $l_{+}^{(6)}$ still occurs in the higher-temperature regime and should therefore not be identified directly with the lower specific-heat crossover. However, unlike $l_{-}^{(6)}$, this facet remains relevant upon further cooling, Fig.~\ref{fig:facet_plus_minus.2Dtrajectory}, as the $P=+1$ sector becomes increasingly populated. Thus, the two facets capture the trends as we evolve from high to low temperature, typified initially by bond-dominated matter-Majorana correlations in the $P=-1$ sector and, upon further cooling, by the growing importance of the $P=+1$ sector and more spatially extended gauge-dressed Majorana correlations, Fig.~\ref{fig:facetplus_minus.1Dtrajectory}.

\subsection{From magic witnesses to reduced descriptions of the local state}
An important observation from our analysis is that the operator spaces identified from the magic witnesses are not restricted to the temperatures at which the corresponding facets are first violated. Although these spaces are recognized from the structure of the witnesses at the onset of magic, they provide an accurate description of the corresponding rdms throughout the full temperature range considered. 
Thus, the operators selected by the RoM optimization appear to capture the structure of the local thermal state that extends for the entire temperature range of the simulation.\\

There is also an algebraic structure underlying this reduced description. The operator spaces identified for the three, four, and six-site marginals close under the Jordan product Eq.~\eqref{eq:JordanProduct} and form finite-dimensional EJAs \cite{JordanVonNeumannWigner1934, FarautKoranyi1994}. The reduction to these spaces preserves a closed algebraic structure among the relevant local observables, which suggests that the magic witnesses identify a reduced operator description intrinsic to the local thermal states, rather than a set of operators relevant only to the particular facet crossing at which they were obtained. \\

More generally, an interesting question is whether similar symmetry-resolved local operator algebras can be identified in other many-body systems and, more importantly, whether they reflect a more general feature of quantum spin liquids and other constrained many-body systems. In particular, it would be interesting to understand whether such reduced algebraic structures arise naturally from the interplay of local constraints, emergent gauge structure, and lattice symmetries.

\subsection{From the bond algebra to the symmetry-resolved plaquette algebra}
The operator structure identified here has a natural connection to the bond-algebraic formulation of the KHM
developed by Nussinov \emph{et al.} \cite{NussinovOrtiz2009,CobaneraOrtizNussinov2011}. In this approach, the algebra generated by the bond terms provides a description of the Hamiltonian, its conserved flux sectors, and the associated correlation functions. Our perspective is complementary: rather than considering the bond algebra as a structure of the full lattice Hamiltonian, we ask what operator content can be carried
by a local marginal. This naturally leads to the symmetry-resolved plaquette algebra introduced above, describing the symmetry-allowed operator content of the hexagonal marginal.\\

The symmetry-resolved construction also makes clear how the local operator structure should change when one of the symmetries is relaxed. In particular, a time-reversal-breaking perturbation, or more generally, a setting in which the time-reversal constraint is lifted while the remaining symmetries are preserved, should therefore enlarge the accessible local operator space from $10$ to $14$ dimensions.

\subsection{Restricted observables and measurement resources}
A practical consequence of the reduced operator description is that the local magic can be characterized using far fewer observables than would be required for a complete reconstruction of the six-site
state. In particular, the symmetry-resolved plaquette algebra reduces the relevant information to a small set of symmetry-adapted expectation values. When resolved into Pauli measurements, these operators can be accessed using only $21$ local measurement settings, compared with the $3^6=729$ settings associated with the complete
six-site Pauli measurement basis.\\

This is, in some sense, in the spirit of the recent works \cite{Varela2026,Liu2026}, where magic is characterized using restricted sets of Pauli strings. The difference, however, is that in our approach the restricted operator set is not specified \emph{a priori}. Rather, it emerges from the structure of the optimal magic witness itself. Furthermore, the operators we considered were linear combinations and polynomials of Pauli strings, rather than being simple Pauli strings. \\

This connection raises a further question: when can a restricted stabilizer problem remain analytically tractable even in the presence of active Pauli dependencies? Liu \emph{et al.} \cite{Liu2026} identify active dependencies as the source of additional sign constraints that are not captured by the frustration graph alone. This raises a natural question: are there structured classes of active dependencies for which these additional sign constraints can nevertheless be resolved analytically?\\

The four-site operator space encountered previously, provides a concrete class in which such
dependencies can be resolved. In particular, consider a centrally dressed family
\begin{equation}
\mathcal M_S=\{S,O_a,SO_a\}_{a=1}^{m}\;,
\end{equation}
where $S$ commutes with the anti-commuting operators $O_a$, and obey the active dependencies
\begin{equation}
S \;O_a\;(SO_a)=I\;.
\end{equation}
The key point is that these relations are controlled by the same center $S=\pm1$. Therefore, resolving the problem into the two $S$ sectors thus removes the apparent sign ambiguity and reduces the dependent problem to two simpler sector problems.\\

This suggests that the relevant distinction may not be simply whether active dependencies are present, but whether they can be resolved through a small number of central or symmetry sectors. Understanding which broader classes of dependency structures admit such a resolution could be an interesting direction to pursue.

\section{Conclusion}
In this work, we investigated the emergence of local magic in the finite-temperature Kitaev honeycomb model through
it's three, four, and six-site marginals. We find that optimal magic witnesses are highly structured and can be expressed in terms of a small number of physically meaningful local operators. The resulting
reduced stabilizer geometries provide a compact description of the thermal evolution of the local state and identify the boundaries across which magic emerges.\\

A central observation is that the operator spaces, and consequently the feature spaces, identified from the magic witnesses are considerably rich in information. Their projections retain the
three, four, and six-site reduced density matrices with very high fidelity throughout the full temperature range considered, and the same reduced description also captures the behavior of a genuine multipartite entanglement witness. For the six-site marginal, this observation can be made stronger. The relevant operator space forms the symmetry-resolved plaquette algebra and provides an exact description of the local state and its robustness of magic when the corresponding symmetries are realized exactly. The hierarchy of operators within this algebra can furthermore be traced
back to the dominant bond contribution in the modular Hamiltonian, while its closure under the Jordan product provides an additional algebraic structure underlying the reduced description.\\

Our results therefore provide a route from the exponentially large local Pauli space to a compact description determined by the microscopic interactions and symmetries of the many-body system. It will be interesting to determine whether similar symmetry-resolved local operator algebras arise in other quantum spin liquids and constrained many-body systems, and whether their structure can be understood more generally from the interplay of local constraints, emergent gauge structure, and lattice symmetries.\\

From an experimental perspective, recent digital realizations of the Kitaev honeycomb model using reconfigurable neutral-atom arrays \cite{Evered2025} provide a natural platform in which these ideas could be investigated. In particular, the local nature of the operator spaces and magic witnesses identified here suggests a route towards probing the emergence and structure of magic directly from local plaquette observables in such quantum simulators.

Note added in upload:~while preparing this paper for
the arXiv, we noticed a preprint reporting complimentary
numerical results for magic in the ground state of the
Kitaev honeycomb model, as determined from a stabilizer
R´enyi entropy \cite{Lamma-arXiv}.   A second preprint, also addressing 
magic in quantum spin models, including the Kitaev honeycomb model \cite{lyu2026fate}, appeared while preparing 
the revised version of this manuscript.

\begin{acknowledgments}
The authors acknowledge helpful discussions with Weslei Fontana and Matthias Gohlke.
SS is grateful for the hospitality of the University of Bristol, where a part 
of this work was carried out.
This work was supported by the Theory of Quantum Matter Unit, OIST, 
by JSPS KAKENHI Grant No. JP25H01247.
Claude Opus 5.0 and ChatGPT 5.5 were used to identify the specific 
algebra associated with the compact operator 
space, and related literature. 
We acknowledge the use of computational resources of the Scientific Computing section 
of the Research Support Division at the Okinawa Institute of Science and Technology 
Graduate University (OIST).

\end{acknowledgments}


All authors contributed to the conceptualization of the work, 
the interpretation of results, 
and the writing of the manuscript.
All analytic and numerical 
calculations were carried out by 
Snigdh Sabharwal. 
Following time--worn tradition, 
authors are listed alphabetically.


\appendix 
\label{sec:Append}

\section{Numerical details}
\label{Append:Numerics}

The KHM model, Eq.~\ref{eq:H.Kitaev}, analysis we presented in the main text, were carried out for a $24$--site cluster with periodic boundary conditions, illustrated in Fig.~\ref{fig:honeycomb.finite.size}.  The rdms $\rho^{(n)}$, used for calculations, were evaluated for a cluster of $n$ spins on a graph embedded within a finite size cluster.
The specific marginals used to produce the results in the main text are illustrated 
in Fig.~\ref{fig:n.site.cluster}.
Density matrices at finite temperature were evaluated using the method of Thermal Pure Quantum 
states (TPQ) \cite{Imada1986,Hams2000,Iitaka2003,Machida2012,Sugiura2013,Ikeuchi2015,Endo2018}, 
implemented through the package H$\Phi$ \cite{Kawamura2017,Hphi-update} as discussed in \cite{Sabharwal2025characterizing}.
At each temperature, an average 
\begin{eqnarray}
\rho^{(n)} (T) &=& \frac{1}{N_{\sf TPQ}} \sum_{i=1}^{N_{\sf TPQ}} \rho^{(n)}_{i} (T)
\end{eqnarray}
was taken over reduced density matrices $\rho^{(n)}_{i} (T)$ constructed from 
\mbox{$N_{\sf TPQ} = 24$} different realization a of TPQ state.

\begin{table*}[t]
\centering
\setlength{\tabcolsep}{10pt}
\renewcommand{\arraystretch}{1.1}
\begin{tabular}{c|l @{\hspace{1cm}} c|l}
\hline\hline
$\mu$ & $\mathbf{a}_{\mu}$ & $\mu$ & $\mathbf{a}_{\mu}$ \\
\hline
1 & $( 1, 1,-1, 1,-1)$
&
5 & $(-1, 1,-1,-1, 1)$
\\
2 & $( 1, 1, 1, 1, 1)$
&
6 & $(-1, 1, 1,-1,-1)$
\\
3 & $( 1,-1, 1,-1, 1)$
&
7 & $(-1,-1, 1, 1,-1)$
\\
4 & $( 1,-1,-1,-1,-1)$
&
8 & $(-1,-1,-1, 1, 1)$
\\
\hline\hline
\end{tabular}
\caption{Coefficient vectors $\mathbf{a}_{\mu}$ defining the eight facets of the
symmetry-reduced four-site stabilizer polytope, with the entries ordered as
specified in Eq.~\ref{eqappend:four.site.feature.vector} and $b_{\mu}=1$
for all facets. Each row defines an inequality
$l_{\mu}^{(4)}
=b_{\mu}+\mathbf{a}_{\mu}\cdot\mathbf{x}^{(4)}\geq0$.
For example, the first row gives
$l_{1}^{(4)}= 1+\langle S^{(4)}\rangle +\langle \Gamma_{1}^{(4)}\rangle
-\langle \Gamma_{2}^{(4)}\rangle +\langle S^{(4)}\Gamma_{1}^{(4)}\rangle
-\langle S^{(4)}\Gamma_{2}^{(4)}\rangle
\geq0$.}
\label{tab:four.site.facets}
\end{table*}

\section{Facet inequalities for the four site marginal}
\label{Append:Half.plane.four.site}

Recall that the projected stabilizer polytope admits the half-space representation
\begin{equation}
    l_{\mu}^{(4)}
    \left[
        \mathbf{x}^{(4)}(\omega)
    \right]
    =
    b_{\mu}
    +
    \mathbf{a}_{\mu}^{\mathsf T}
    \mathbf{x}^{(4)}(\omega)
    \geq 0.
\end{equation}
For completeness, we use the feature ordering
\begin{align}
\label{eqappend:four.site.feature.vector}
    \mathbf{x}^{(4)}(\omega)
    =
    \big(
    &\langle S\rangle_{\omega},
    \langle \Gamma_1\rangle_{\omega},
    \langle \Gamma_2\rangle_{\omega}, \langle S\Gamma_1\rangle_{\omega},
    \langle S\Gamma_2\rangle_{\omega},
    \big).
\end{align}

Each facet is therefore specified by the constant term $b_{\mu}$ and the
five-dimensional coefficient vector $\mathbf{a}_{\mu}$. In this feature
space, the projected stabilizer polytope has $8$ facets. For all of them,
the constant term is
\begin{equation}
\label{eq:bmu.four.site}
    b_{\mu}=1,
    \qquad
    \mu=1\;,\ldots,8.
\end{equation}
The corresponding coefficient vectors $\mathbf{a}_{\mu}$ are listed in
Table~\ref{tab:four.site.facets}.

\section{Facet inequalities for the six-site marginal}\label{Append:Half.plane.six.site}
Here we collect the facet inequalities relevant for the six-site marginal.
Recall that the stabilizer polytope can be written in its half-space
representation as
\begin{equation}
    l_{\mu}^{(6)}
    \left[
        \mathbf{x}^{(6)}(\omega)
    \right]
    =
    b_{\mu}
    +
    \mathbf{a}_{\mu}^{\mathsf T}
    \mathbf{x}^{(6)}(\omega)
    \geq 0 .
\end{equation}
For completeness, the feature vector is ordered as
\begin{align}\label{eqappend:six.site.feature.vector}
\mathbf{x}^{(6)}(\omega)
=
\big(
&\langle P\rangle_{\omega},
\langle B\rangle_{\omega},
\langle C_1\rangle_{\omega},
\langle C_2\rangle_{\omega},
\langle D_1\rangle_{\omega},
\langle D_2\rangle_{\omega},
\nonumber\\
&\langle PB\rangle_{\omega},
\langle PC_1\rangle_{\omega},
\langle PC_2\rangle_{\omega}
\big).
\end{align}
The resulting $18$ facet inequalities are specified by the
coefficients in Table~\ref{tab:six.site.facets}.
\begin{table*}[t!]

\centering

\scriptsize

\setlength{\tabcolsep}{10pt}

\renewcommand{\arraystretch}{1.05}

\resizebox{\textwidth}{!}{%
\begin{tabular}{c|c|l @{\hspace{0.5cm}} c|c|l}

\hline\hline

$\mu$ & $b_{\mu}$ & $\mathbf{a}_{\mu}$
&
$\mu$ & $b_{\mu}$ & $\mathbf{a}_{\mu}$
\\

\hline

\multicolumn{6}{c}{\textit{Trivial facets}}\\
\hline

1 & $1$ &
$(1,0,0,-1,-1,1,0,0,-1)$
&
5 & $1$ &
$(1,0,0,-1,1,-1,0,0,-1)$
\\

2 & $3$ &
$(3,2,2,1,1,3,2,2,1)$
&
6 & $3$ &
$(3,1,-1,1,-1,-3,1,-1,1)$
\\

3 & $3$ &
$(3,-1,-1,1,1,3,-1,-1,1)$
&
7 & $3$ &
$(-3,0,-1,-1,0,0,0,1,1)$
\\

4 & $3$ &
$(3,-2,2,1,-1,-3,-2,2,1)$
&
& &
\\

\hline
\multicolumn{6}{c}{\textit{Non-trivial facets}}\\
\hline

8 & $3$ &
$(3,1,-1,-1,1,-3,1,-1,-1)$
&
14 & $3$ &
$(-3,2,1,1,0,0,-2,-1,-1)$
\\

9 & $3$ &
$(3,-1,-1,-1,-1,3,-1,-1,-1)$
&
15 & $3$ &
$(-3,1,2,-1,0,0,-1,-2,1)$
\\

10 & $3$ &
$(-3,0,-1,1,0,0,0,1,-1)$
&
16 & $3$ &
$(-3,-1,0,3,0,0,1,0,-3)$
\\

11 & $3$ &
$(-3,-1,2,-1,0,0,1,-2,1)$
&
17 & $3$ &
$(-3,-1,0,-1,0,0,1,0,1)$
\\

12 & $3$ &
$(-3,1,0,-1,0,0,-1,0,1)$
&
18 & $3$ &
$(-3,-2,1,1,0,0,2,-1,-1)$
\\

13 & $3$ &
$(-3,1,0,3,0,0,-1,0,-3)$
&
& &
\\

\hline\hline

\end{tabular}%
}

\caption{
Coefficients $b_{\mu}$ and $\mathbf{a}_{\mu}$ defining the $18$ facets
of the symmetry-reduced six-site stabilizer polytope, with the entries of
$\mathbf{a}_{\mu}$ ordered as specified in
Eq.~\ref{eqappend:six.site.feature.vector}. The polytope contains
$15$ vertices and $18$ facets, of which $\mu=1,\ldots,7$ are trivial
facets and $\mu=8,\ldots,18$ are the $11$ non-trivial facets.
Each row defines an inequality
$l_{\mu}^{(6)}
=b_{\mu}+\mathbf{a}_{\mu}\cdot\mathbf{x}^{(6)}\geq0$.
}
\label{tab:six.site.facets}
\end{table*}
\section{Euclidean Jordan algebra for the six-site marginal}\label{Append:sec.algebra}
Here we show explicitly that the ten-dimensional operator space introduced
for the six-site marginal,
\begin{equation}
\mathcal{V}^{(6)} = \text{span}_{\mathbb{R}} \left\{I,P,B,C_1,C_2,D_1,D_2,PB,PC_1,PC_2 \right\},
\end{equation}
forms a Euclidean Jordan algebra (EJA). The nontrivial Jordan products involving the plaquette operator are
\begin{subequations}   
\begin{align}
P\circ P &= I,\\
P\circ B &= PB,\\
P\circ C_1 &= PC_1,\\
P\circ C_2 &= PC_2,\\
P\circ D_1 &= D_1,\\
P\circ D_2 &= D_2,\\
P\circ PB &= B,\\
P\circ PC_1 &= C_1,\\
P\circ PC_2 &= C_2.
\end{align}
\end{subequations}   
The remaining nontrivial products between the undressed basis operators are
\begin{subequations}   
\begin{align}
B\circ B &= 6I+2C_1+2C_2,\\
B\circ C_1 &= 2B+D_1+3D_2,\\
B\circ C_2 &= B,\\
B\circ D_1 &= C_1+PC_1,\\
B\circ D_2 &= C_1+PC_1,\\
C_1\circ C_1 &= 6I+2C_1+2PC_2,\\
C_1\circ C_2 &= PC_1,\\
C_1\circ D_1 &= B+PB,\\
C_1\circ D_2 &= B+PB,\\
C_2\circ C_2 &= 3I-2PC_2,\\
C_2\circ D_1 &= -2D_1+3D_2,\\
C_2\circ D_2 &= D_1,\\
D_1\circ D_1 &= 6I+6P-4C_2-4PC_2,\\
D_1\circ D_2 &= 2C_2+2PC_2,\\
D_2\circ D_2 &= 2I+2P.
\end{align}
\end{subequations}
Since $P$ commutes with every element of $\mathcal{V}^{(6)}$ and $P^2=I$, the products involving $PB$, $PC_1$, and $PC_2$ follow from
\begin{align}
(P M)\circ N &= P(M\circ N),\\
(P M)\circ(P N) &= M\circ N,
\end{align}
with $M,N \in {B, C_1, C_2}$. These relations show explicitly that $\mathcal{V}^{(6)}$ is closed under the
Jordan product, and therefore forms a ten-dimensional EJA.

\section{Majorana representation of the six-site operator space}
\label{Append:Majorana.operator.space}
Here we go over the Majorana representation of the operators spanning the six-site operator space, $\mathcal{V}^{(6)}$. Following \cite{Kitaev2006}, we write
\begin{equation}
    \sigma_i^{\alpha}=i b_i^{\alpha}c_i,
\end{equation}
where $c_i$ denotes the matter (itinerant) Majorana fermion and $b_i^{\alpha}$ the gauge Majorana associated with spin component $\alpha$. For an $\alpha$-type bond $\langle ij\rangle_{\alpha}$, we introduce the $\mathbb{Z}_2$ link operator
\begin{equation}
    u_{ij}^{(\alpha)}=i b_i^\alpha b_j^\alpha,
    \qquad
    i\in \Lambda_A,\quad j\in \Lambda_B,
\end{equation}
such that
\begin{equation}
    \sigma_i^\alpha\sigma_j^\alpha
    =
    -i u_{ij}c_ic_j.
\end{equation}
For the six-site marginal, we label the sites cyclically by $1\ldots6$ with
\begin{equation}
    1, 3, 5 \in \Lambda_{A}, \qquad 2,4, 6 \in \Lambda_B
\end{equation}
We define the gauge links according to the fixed $\Lambda_A \to \Lambda_B$ orientation,
\begin{equation}
\begin{aligned}
    u_1&\equiv u_{12}^{(z)},&
    u_2&\equiv u_{32}^{(y)},&
    u_3&\equiv u_{34}^{(x)},\\
    u_4&\equiv u_{54}^{(z)},&
    u_5&\equiv u_{56}^{(y)},&
    u_6&\equiv u_{16}^{(x)}.
\end{aligned}
\label{eq:Majorana.link.orientation}
\end{equation}
Consequently, the bond operators become
\begin{equation}
\begin{aligned}
    K_1&=-iu_1c_1c_2,&
    K_2&=+iu_2c_2c_3,&
    K_3&=-iu_3c_3c_4,\\
    K_4&=+iu_4c_4c_5,&
    K_5&=-iu_5c_5c_6,&
    K_6&=+iu_6c_6c_1.
\end{aligned}
\label{eq:Majorana.K}
\end{equation}
Their sum defines the nearest-neighbor bond operator,
\begin{align}
B=&K_1 + K_2 + K_3 + K_4 + K_5 + K_6\\
=&-i\big(
u_1c_1c_2-u_2c_2c_3+u_3c_3c_4-u_4c_4c_5\nonumber\\
&+u_5c_5c_6-u_6c_6c_1
\big).
\end{align}
Hence, $B$ contains the nearest-neighbor matter-Majorana bilinears entering the bond structure.

Next, we consider the plaquette operator, which, recall, is the product of the six bond operators,
\begin{equation}
    P=K_1K_2K_3K_4K_5K_6\;.
\end{equation}
Substituting Eq.~\eqref{eq:Majorana.K}, all matter Majoranas cancel and we obtain
\begin{equation} \label{eq:Majorana.P}
    P=u_1u_2u_3u_4u_5u_6\;.
\end{equation}
Thus, $P$ specifies the local $\mathbb{Z}_2$ flux through the hexagon.
Likewise, the first two-bond orbit sum is
\begin{align}
    C_1 =&K_1K_3+K_2K_4+K_3K_5 +K_4K_6+K_5K_1+K_6K_2\\
        =& -\big( u_1u_3\,c_1c_2c_3c_4 +u_2u_4\,c_2c_3c_4c_5 +u_3u_5\,c_3c_4c_5c_6 \nonumber\\
                &+u_4u_6\,c_4c_5c_6c_1 +u_5u_1\,c_5c_6c_1c_2 +u_6u_2\,c_6c_1c_2c_3 \big).
\end{align}
The second two-bond orbit sum is
\begin{align}
    C_2  =& K_1K_4 + K_2K_5 + K_3K_6 \\
        =& u_1u_4\,c_1c_2c_4c_5 +u_2u_5\,c_2c_3c_5c_6+u_3u_6\,c_3c_4c_6c_1\;.
\end{align}
Thus, $C_1$ and $C_2$ correspond to gauge-dressed four matter-Majorana correlations generated by products of two bonds.

The first three-bond orbit sum is
\begin{align}
    D_1 &=-(K_1K_2K_3 +K_2K_3K_4 +K_3K_4K_5 \nonumber\\
        &\quad\;+K_4K_5K_6+K_5K_6K_1+K_6K_1K_2) \\
        = & i\bigg[(u_1u_2u_3+u_4u_5u_6)c_1c_4 - (u_2u_3u_4+u_5u_6u_1)c_2c_5\nonumber \\
        &+ (u_3u_4u_5+u_6u_1u_2)c_3c_6 \bigg].
\end{align} 
Thus, $D_1$, which is generated by products of three bond operators, reduces to a sum of matter-Majorana bilinears connecting opposite sites of the hexagon, dressed by three-link gauge strings. The second three-bond orbit sum is
\begin{align}
    D_2 =&K_1K_3K_5+K_2K_4K_6 \\
        =&i\left(u_1u_3u_5+u_2u_4u_6 \right) c_1c_2c_3c_4c_5c_6.
\end{align}
Thus, $D_2$ corresponds to a gauge-dressed six-Majorana structure associated with the two alternating bond configurations.

The remaining independent operators are obtained by dressing $B$, $C_1$, and $C_2$ with the plaquette operator $P$, while $D_1$ and $D_2$ are already invariant under multiplication by $P$.

For the plaquette-dressed bond operator, we obtain
\begin{equation}
\begin{aligned}
PB=-i\big(
&u_2u_3u_4u_5u_6\,c_1c_2
-u_1u_3u_4u_5u_6\,c_2c_3\\
&+u_1u_2u_4u_5u_6\,c_3c_4
-u_1u_2u_3u_5u_6\,c_4c_5\\
&+u_1u_2u_3u_4u_6\,c_5c_6
-u_1u_2u_3u_4u_5\,c_6c_1
\big).
\end{aligned}
\label{eq:Majorana.PB}
\end{equation}
Similarly, for $PC_1$ we obtain
\begin{equation}
\begin{aligned}
PC_1=-\big(
&u_2u_4u_5u_6\,c_1c_2c_3c_4
+u_1u_3u_5u_6\,c_2c_3c_4c_5\\
&+u_1u_2u_4u_6\,c_3c_4c_5c_6
+u_1u_2u_3u_5\,c_4c_5c_6c_1\\
&+u_2u_3u_4u_6\,c_5c_6c_1c_2
+u_1u_3u_4u_5\,c_6c_1c_2c_3
\big).
\end{aligned}
\label{eq:Majorana.PC1}
\end{equation}
Finally, $PC_2$ takes the form
\begin{equation}
\begin{aligned}
PC_2={}&
u_2u_3u_5u_6\,c_1c_2c_4c_5
+u_1u_3u_4u_6\,c_2c_3c_5c_6\\
&+u_1u_2u_4u_5\,c_3c_4c_6c_1.
\end{aligned}
\label{eq:Majorana.PC2}
\end{equation}

The Majorana representation therefore gives a complementary interpretation of the operator hierarchy appearing in the six-site marginal. The plaquette operator $P$ specifies the local $\mathbb{Z}_2$ flux, while $B$ contains nearest-neighbor matter-Majorana bilinears. The operators $C_1$ and $C_2$ describe gauge-dressed four-Majorana correlations. Interestingly, $D_1$ reduces again to a matter-Majorana bilinear structure connecting opposite sites of the hexagon through three-link gauge strings, whereas $D_2$ contains a gauge-dressed six-Majorana structure associated with the two alternating bond configurations. The operators $PB$, $PC_1$, and $PC_2$ provide the corresponding matter-Majorana correlations resolved with respect to the plaquette-flux sector.

\bibliographystyle{apsrev4-2}
\bibliography{main_with_C1_C2_D1_D2}

\end{document}